\documentclass[pdftex,twocolumn,epjc3]{svjour3}     

\RequirePackage[T1]{fontenc}

\smartqed  

\RequirePackage{graphicx}
\RequirePackage{mathptmx}      
\RequirePackage{flushend}
\RequirePackage[numbers,sort&compress]{natbib}
\RequirePackage[colorlinks,citecolor=blue,urlcolor=blue,linkcolor=blue]{hyperref}
\usepackage{amsmath}
\usepackage{amsfonts}
\usepackage{mathrsfs}
\usepackage{amssymb}
\usepackage{subfigure}
\usepackage{multicol}
\usepackage{tabularx}
\usepackage{feynmp}
\usepackage{multirow}
\usepackage{slashed}
\usepackage{units}
\usepackage{bbm}
\usepackage{framed}
\usepackage[normalem]{ulem}
\usepackage{mathtools}
    \usepackage{makeidx}
    \usepackage{subfigure}
    \usepackage{booktabs}
    \usepackage{tabu}
    \usepackage[thinlines]{easytable}
	\usepackage{MnSymbol}
    \usepackage{array}
\usepackage{hyphenat}

\newcommand{\be}{\begin{eqnarray}}
\newcommand{\ee}{\end{eqnarray}}
\def\nn{\nonumber}

\DeclareMathOperator*{\SumInt}{%
\mathchoice%
  {\ooalign{$\displaystyle\sum$\cr\hidewidth$\displaystyle\int$\hidewidth\cr}}
  {\ooalign{\raisebox{.14\height}{\scalebox{.7}{$\textstyle\sum$}}\cr\hidewidth$\textstyle\int$\hidewidth\cr}}
  {\ooalign{\raisebox{.2\height}{\scalebox{.6}{$\scriptstyle\sum$}}\cr$\scriptstyle\int$\cr}}
  {\ooalign{\raisebox{.2\height}{\scalebox{.6}{$\scriptstyle\sum$}}\cr$\scriptstyle\int$\cr}}
}
\begin{document}

\title{Dynamical symmetry breaking, magnetization and induced charge in graphene: Interplay between magnetic and pseudomagnetic fields}


\author{J. A. S\'anchez-Monroy,\thanksref{e1,addr1,addr2}
        \and
        C. J. Quimbay\thanksref{e2,addr1} 
}


\thankstext{e1}{e-mail: jasanchezm@unal.edu.co}
\thankstext{e2}{e-mail: cjquimbayh@unal.edu.co}

\institute{Departamento de F\'{\i}sica, Universidad Nacional de Colombia,
Bogot\'a, D. C., Colombia\label{addr1}
\and 
Instituto de F\'{\i}sica, Universidade de S\~{a}o Paulo, 05508-090, S\~{a}o Paulo, SP, Brazil\label{addr2}
}


\maketitle

\begin{abstract}
In this paper, we investigate the two competing effects of strains and magnetic fields in single-layer graphene to explore its impact on various phenomena of quantum field theory, such as induced charge density, magnetic catalysis, symmetry breaking, dynamical mass generation and magnetization. We show that the interplay between strains and magnetic fields produces not only a breaking of chiral symmetry, as it happens in QED$_{2+1}$, but also parity and time-reversal symmetry breaking. The last two symmetry breakings are related to the dynamical generation of a Haldane mass term. We find that it is possible to modify the magnetization and the dynamical mass independently for each valley, by strain and varying the external magnetic field. Furthermore, we discover that the presence of a non-zero pseudomagnetic field, unlike the magnetic one, allows us to observe an induced ``vacuum'' charge and a parity anomaly in strained graphene. Finally, because the combined effect of real and pseudomagnetic fields produces an induced valley polarization, the results presented here may provide new tools to design valleytronic devices.
\end{abstract}

\section{Introduction}
In recent times, a number of materials for which quasiparticle excitations behave like relativistic two-dimensional fermions have appeared in condensed matter. One of the most fascinating examples of such materials is single-layer graphene, which is a material consisting of a single-layer of graphite. Graphene exhibits many interesting features, among which are the anomalous quantum Hall effect \cite{Zhang2005Exp}, a record high Young's modulus \cite{Lee2008Ela}, ultrahigh electron mobility \cite{Bolotin2008}, as well as very high thermal conductivity \cite{Faugeras2010a}. The band structure of single-layer graphene has two inequivalent and degenerate valleys, $\vec{K}$ and $\vec{K}'$, at opposite corners of the Brillouin zone. The possibility to manipulate the valley to store and carry information defines the field of ``valleytronics'', in a analogous way as the role played by spin in spintronics.
\par
One of the most exciting aspects of the physics of single-layer graphene is that several unobservable phenomena in experiments of high energy physics may be observed, such as Klein tunneling \cite{katsnelson2006chiral} and ‘‘Zitterbewegung’’ \cite{KATSNELSON20073}. From the point of view of quantum field theory, graphene exhibits similar features with quantum electrodynamics in three dimensions (QED$_{2+1}$)\footnote{Since the electrostatic potential between two electrons on a plane is the usual $1/r$ Coulomb potential instead of a logarithmic potential, which is distinctive of quantum electrodynamics in $2+1$ dimensions (QED$_{2+1}$),
the theory that describes graphene at low energies is known as \textit{reduced quantum electrodynamics} (RQED$_{4,3}$) \cite{Marino1993a,Gorbar2001,Teber2012a}. In RQED$_{4,3}$, the fermions are confined to a plane; nevertheless, the electromagnetic interaction between them is three-dimensional.} in such a way that it is possible to explore sophisticated aspects of three-dimensional quantum field theory; for instance, magnetic catalysis, symmetry breaking, dynamical mass generation, and anomalies, among others. It is possible, in the first place, because the two valleys can be associated with two irreducible representations of the Clifford algebra in three dimensions. In the second place, because the low-energy regime quasiparticles behave like massless relativistic fermions, where the speed of light is replaced by the Fermi velocity $v_F$, which is about 300 times smaller than the speed of light.
\par
The presence of a mass gap may turn single-layer graphene from a semimetal into a semiconductor. This can be accomplished, for example, when a single-layer graphene sheet is placed on a hexagonal boron nitride substrate \cite{Giovannetti2007,Hunt2013}, 
or deposited on a SiO$_2$ surface \cite{Shemella2009}. Additionally, it has been proposed that a bandgap can be induced by vacuum fluctuations \cite{Kibis2011}. Significantly, a mass gap suppresses the Klein tunneling so that this fact could be useful in the design of devices based on single-layer graphene \cite{Navarro2018}.
\par
The phenomenon known as \textit{magnetic catalysis} appears when a dynamical symmetry breaking occurs in the presence of an external magnetic field, independent of its intensity \cite{Gusynin1994,Shovkovy2013}. Since in QED$_{2+1}$ the mass term breaks the chiral symmetry in a reducible representation\footnote{Because the chiral symmetry cannot be defined for irreducible representations, it does not make sense to talk about chiral symmetry breaking.}, magnetic catalysis rise as $m\rightarrow 0$. Dynamical symmetry breaking is a consequence of the appearance of a nonvanishing chiral condensate $\langle 0|T[\Psi,\bar{\Psi}]|0\rangle$, which leads to the generation of a fermion dynamical mass \cite{Gusynin1994}. In particular, when single-layer graphene is subjected to an external magnetic field, a nonvanishing chiral condensate ensures that there will be a dynamical chiral symmetry breaking \cite{Gusynin1994}, as well as a dynamical mass equal for each valley \cite{Farakos1998b,Raya2010}.
\par
When a sample of single-layer graphene presents strains, ripples or curvature, the dispersion relation is modified in such a way that an effective gauge vector field coupling is induced in the low energy Dirac spectrum (the so-called \textit{pseudomagnetic field} \cite{Guinea2010a}).  The mechanical control over the electronic structure of graphene has been explored as a potential approach to \textit{``strain engineering''} \cite{Levy2010,Zhu2015a}. Originally, it was observed that strain produces a strong gauge field that effectively acts as a uniform pseudomagnetic field whose intensity is greater than $10$ T \cite{Guinea2010}, a pseudomagnetic field greater than $300$ T was experimentally reported later \cite{Levy2010}. This pseudomagnetic field opens the door to previously inaccessible high magnetic field regimes.
\par
In contrast to the case of a real external magnetic field, the pseudomagnetic field experienced by the particles in the valleys $\vec{K}$ and $\vec{K}'$ have opposite signs. Hence, when a sample of strain single-layer graphene is placed in a perpendicular magnetic field, the energy levels suffer a different separation for each valley, which results in an induced valley polarization \cite{Tony2010}. The previous one is precisely the key requirement for valleytronic devices. Beyond theoretical calculations, the presence of Landau Levels in graphene has been experimentally observed in external magnetic fields \cite{Jiang2007a}, strain-induced pseudomagnetic fields \cite{Levy2010,Yan2012a} and in the coexistence of pseudomagnetic fields and external magnetic fields \cite{Li2015a}. Moreover, the effects of the combination of an external magnetic field and a strain-induced pseudomagnetic field in different configurations were studied in order to construct a valley filter \cite{Feng2010a,PChaves2010a,Fujita2010a,Feng2011a}.
\par
In the first part of this paper, we study how an interplay between real and pseudomagnetic fields affects the symmetry breaking and the dynamical mass generation. As we will see, the presence of these two fields produces not only a breaking of chiral symmetry but also parity and time-reversal symmetry breaking. Furthermore, we will show that there will be a dynamical mass generation of two types, the usual mass ($m\bar{\psi}\psi$) and another known as Haldane mass \cite{Haldane1988a}, unlike QED$_{2+1}$ where only the usual mass term is dynamically generated. As a result of this, the dynamical fermion masses will be different for each valley. In this paper, we will use a non-perturbative method based on the quantized solutions of the Dirac equation, the so-called \textit{Furry picture}. 
The reason for using this method is to obtain nonperturbative results since the effective coupling constant in graphene is of order unity, $\alpha \approx 2.5$, raising serious questions about the validity of the perturbation expansion in graphene \cite{Kolomeisk2015a}. However, the latter has been a matter of controversy, since at low energy it was experimentally observed that the effective fine-structure constant approaches $1/7$ \cite{Reed2010}.
\par
In the second part, we investigate how the presence of real and pseudomagnetic fields affects the magnetization. In conventional metals, the magnetism receives contributions from spin (Pauli paramagnetism) and orbitals (Landau diamagnetism). In particular, the orbital magnetization of graphene in a magnetic field has shown a non-linear behavior as a function of the applied field \cite{Slizovskiy2012}. In order to examine how the magnetization and the susceptibility behave for each valley in the presence of constant magnetic and pseudomagnetic fields, we will first obtain the one-loop effective action. The one-loop effective action without strains in $2+1$ dimensions had been previously calculated within the Schwinger's proper time formalism \cite{Redlich1984a,Andersen1995a,Dittrich1997} and using the fermion propagator expanded over the Landau levels \cite{Ayala2010c}. Taking into account that in static background fields the one-loop effective action is proportional to the vacuum energy \cite{Weinberg1996}, which can be calculated in a direct way employing the furry picture, we use this method to study the most general case. As we will show, the presence of magnetic and pseudomagnetic fields allows us to manipulate the magnetization and the susceptibility of each valley independently.
\par
Finally, we study the parity anomaly and the induced vacuum charge in strained single-layer graphene. In quantum field theory, if a classical symmetry is not conserved at a quantum level, it is then said that the theory suffers from an anomaly. For instance, in QED$_{2+1}$ if one maintains an invariant gauge regularization in all the calculations, with an odd number of fermion species, the parity symmetry is not preserved by quantum corrections, \textit{i.e.} it has a parity anomaly \cite{Semenoff1983a,Semenoff1984a,Redlich1984a}. In this way, the quantum correction to the vacuum expectation value of the current can be computed to characterize the parity anomaly. As it was pointed out by Semenoff \cite{Semenoff1984a}, external magnetic fields induce a current ($j_{\mu}$) of abnormal parity in the vacuum for each fermion species. Unfortunately, for a even number of fermion species, the total current is canceled: $J_+^{\mu}=j_{1}^{\mu}+j_{2}^{\mu}=0$, and therefore the induced vacuum current is not directly observable. Therefore, it is possible to maintain the gauge and parity symmetries even at the quantum level \cite{Redlich1984a}. In the literature, a number of scenarios were proposed to realize parity anomaly in $2+1$ dimensions. Haldane introduces a condensed matter lattice model in which parity anomaly takes place when the parameters reach critical values \cite{Haldane1988a}. Obispo and Hott \cite{Obispo2014, Obispo2015} show that graphene coupling to an axial-vector gauge potentially exhibits parity anomaly and fermion charge fractionalization. Zhang and Qiu \cite{Zhang2017a} show that in a graphene-like system, with a finite bare mass, a parity anomaly related $\rho$-exciton can be generated by absorbing a specific photon. Alternatively, as Semenoff remarked \cite{Semenoff1984a}, one could consider an ``unphysical'' field with abnormal parity coupled to the fermions, since for such field the total induced vacuum current should be different from zero and hence observable. As we will show, this field is physical, and it is just a pseudomagnetic field with a simple uniform field profile.
\par
This paper is organized as follows: In section \ref{secDiracHamiltonian}, we introduce the Dirac Hamiltonian and the symmetries for single-layer graphene in a finite mass gap. In section \ref{secFurry}, we present the Furry picture for fermions in the presence of real and pseudomagnetic fields.  In section \ref{secCondesateM}, we compute the magnetic condensate and discuss how this characterizes the symmetry breaking and its connection with the dynamical mass of each valley. In section \ref{oneloopeamagnetization}, the one-loop effective action and the magnetization are calculate for each valley. In section \ref{secIndChar}, we calculate the total induced vacuum charge density to show that graphene in the presence of a pseudomagnetic field exhibits a parity anomaly. In appendix \ref{ap:ConsRPMF}, we obtain the exact solution of the Dirac equation for uniform real and pseudomagnetic fields. In appendix \ref{ap:MagConApen}, we compare the calculation of fermionic condensate in the Furry picture with the method via fermion propagator and prove that the trace of the fermion propagator evaluated at equal space-time points must be understood as the expectation value of the commutator of two field operators. Finally, section \ref{conclutionsfinal} contains our conclusions.
\section{Dirac Hamiltonian for graphene}\label{secDiracHamiltonian}
In a vicinity of the Fermi points, the Dirac Hamiltonian in the presence of real ($A$) and pseudo ($a$) magnetic potentials reads ($\hbar=v_F=1$) \cite{Herbu2008a,Kim2011a}\footnote{The electric charge $e$ is multiplying the term $a_i$ only for dimensional reasons; strictly speaking, one could write $eA_i=\tilde{A}_i$ to emphasize that it is independent of $e$.}
\be\label{DHFourC}
H_D[A,a]=\Gamma^0\Gamma^i(p_i+eA_i+iea_i)+\Gamma^0 m,
\ee
where $m$ is a mass gap, $a_i=a_i^{35}\Gamma^{35}$, $\Gamma^{35}=i\Gamma^{3}\Gamma^{5}$\footnote{It turns out that one can identify $a_i^{35}$ as one component of a non-Abelian $SU(2)$ gauge field within the low-energy theory of graphene \cite{Roy2011,Gopalakrishnan2012a}. The other two components of this non-Abelian $SU(2)$ gauge field are proportional to $\Gamma^3$ and $\Gamma^5$, since they are off-diagonal in valley index mixing the two inequivalent valleys \cite{Roy2011,Gopalakrishnan2012a}. In this case, the pseudo-gauge potential is $a_i=a_i^3\Gamma^3 + a_i^5\Gamma^5 +a_i^{35}\Gamma^{35}$. Assuming a smooth enough deformation in the graphene sheet, one can keep only the component $a_i^{35}$, which does not mix the two inequivalent valleys \cite{Roy2011}. Hence, Eq. (\ref{DHFourC}) captured the physics of low-energy strained graphene.}. This $4\times 4$ Hamiltonian acts on the  four-component ``spinor'', $\psi^T=(\psi^K_A,\psi^K_B,\psi^{K'}_A,\psi^{K'}_B)$, where the components take into account both two \textit{valleys} ($\vec{K}$ and $\vec{K}'$) and the two \textit{sublattices} (A and B) \cite{Katsnelson2012book}, the quantum number associated with the two sublattices is usually referred to as \textit{pseudospin}. If one wants to include the real spin, the spinor will have eight-components, and the Dirac Hamiltonian will be $H_{D(8\times 8)}=I_2\otimes H_D[A,a]$ \cite{Roy2011}. For subsequent calculations, it is sufficient to consider $H_D[A,a]$, given that including the real spin only increases the degeneration of the Landau levels by two ($g_s=2$). Since the difference between QED$_{2+1}$ and RQED$_{4,3}$ lies in the kinetic term of the gauge fields, and the magnetic field here is considered as an external field and the pseudomagnetic is a non-dynamical field. Then Eq. (\ref{DHFourC}) is an appropriate description for strained graphene in the presence of an external magnetic field.
\par
As a matter of convenience, we choose here the $\Gamma-$ matrices as \cite{Gomes2009a}
\begin{align}\nn
\Gamma^0&=\sigma^3\otimes \sigma^3=\left(\begin{array}{cc}
\sigma^3 & 0 \\
0 & -\sigma^3
\end{array}\right), \\\nn
\Gamma^1&=\sigma^3\otimes i\sigma^1=\left(\begin{array}{cc}
i\sigma^1 & 0 \\
0 & -i\sigma^1
\end{array}\right),
\\\nn
\Gamma^2&=\sigma^3\otimes i\sigma^2=\left(\begin{array}{cc}
i\sigma^2 & 0 \\
0 & -i\sigma^2
\end{array}\right),\\\nn
\Gamma^3&=i\sigma^1\otimes I_{2\times 2}=\begin{pmatrix}
0  & \ iI \\
iI  & \ 0
\end{pmatrix},
\\\nn
\Gamma^5&=-\sigma^2\otimes I_{2\times 2}=\left(\begin{array}{cc}
0 & \ iI \\
-iI & \ 0
\end{array}\right),
\\
\Gamma^{35}&=i\sigma^3\otimes I_{2\times 2}=\left(\begin{array}{cc}
iI & \ 0 \\
0 & \ -iI
\end{array}\right),
\end{align}
so that $(\Gamma^{3})^2=-1$, $(\Gamma^{5})^2=1$, $\Gamma^{3}$ and $\Gamma^{5}$ anticommute with $\Gamma^{\mu}$, while $\Gamma^{35}$ commutes with $\Gamma^{\mu}$ and anticommutes with $\Gamma^{3}$ and $\Gamma^{5}$. Note that $\Gamma^{\mu}$ ($\mu=0,1,2$) are block-diagonal, where each block  is one of two inequivalent irreducible representations of the Clifford algebra in $2+1$ dimensions. For odd dimensions, there are two inequivalent irreducible representations of the Dirac matrices that we denote as $\mathcal{R}_1$ and $\mathcal{R}_2$. The two inequivalent representations were chosen as $\gamma^{\mu}$ and $-\gamma^{\mu}$ for $\mathcal{R}_1$ and $\mathcal{R}_2$, respectively, where
\begin{equation}
\gamma^{0}=\sigma^3, \ \  \gamma^{1}=i\sigma^1,\ \
\gamma^{2}=
i\sigma^2.
\end{equation}
Given that there is no intervalley coupling, we can rewrite the Dirac Hamiltonian as $H_D=H_+[A,a]\oplus H_-[A,a]$, thus	
\be\label{DHtwovalleys}
H_{\pm}=i\sigma^3\sigma^i(p_i+eA_i\mp ea_i^{35})\pm\sigma^3 m,
\ee
where $H_+$ and $H_-$ represent the Hamiltonian near of valley $\vec{K}$ (representation $\mathcal{R}_1$) and $\vec{K}'$ (representation $\mathcal{R}_2$), respectively. $H_+$ acts in a two-component spinor that describes a fermion with pseudospin up and an antifermion with pseudospin down, while $H_-$ acts in a two-component spinor that describes a fermion with pseudospin down and an antifermion with pseudospin up. Thus, we obtain two decoupled Dirac equations in $(2+1)$-dimensions
\be\label{DiracEpm}
i\frac{\partial \psi(x,y,t)}{\partial t}=H_{\pm}\psi(x,y,t).
\ee
Finally, we can write the Lagrangian density for this system as the sum of two Lagrangian densities for each valley
\be\nn
\mathcal{L}=\overbrace{i\bar{\psi}^K\gamma^{\mu}D^+_{\mu}\psi^K-m\bar{\psi}^{K}\psi^{K}}^{\mathcal{L}_{+}}+
\overbrace{i\bar{\psi}^{K'}\gamma^{\mu}D^-_{\mu}\psi^{K'}+m\bar{\psi}^{K'}\psi^{K'}}^{\mathcal{L}_{-}},\\\label{DiracLagratwovalleys+-}
\ee
where $D^{\pm}_{i}=\partial_{i}+ieA_i\mp iea_i^{35}$, which can be interpreted as a system describing two species of two-component spinors, one with mass $+m$ and coupled to $eA_i-ea_i^{35}$ and the other with mass $-m$ and coupled to $eA_i+ea_i^{35}$. Hence, in the vicinity of the Fermi points, graphene monolayers constitute an ideal scenario to simulate the matter sector of QED$_{2+1}$. In the following, we will neglect the corrections due to the effects of Coulomb interactions between the charge carriers. However, we point out that the model given by Eq. (\ref{DHFourC}) is in good agreement with the experiments carried out in graphene in the presence of external magnetic fields \cite{Jiang2007a,novoselov2007rise}, pseudomagnetic fields \cite{Levy2010, Yan2012a}, and in the combination of magnetic and pseudomagnetic fields \cite{Li2015a}.

\subsection{Symmetries in the irreducible and reducible representations}\label{SymIrreRe}
\textbf{Irreducible representations:} For irreducible representations, it is possible to define the parity ($\mathcal{P}$), charge conjugation ($\mathcal{C}$), and time-reversal ($\mathcal{T}$) transformations as follows:
\begin{align}
\mathcal{P}\psi(t,x,y)\mathcal{P}^{-1}&=-i\gamma^1\psi(t,-x,y),\\
\mathcal{T}\psi(t,x,y)\mathcal{T}^{-1}&=-i\gamma^2\psi(-t,x,y),\\\label{Ccharge2x2}
\mathcal{C}\psi(t,x,y)\mathcal{C}^{-1}&=-\gamma^2(\bar{\psi}(t,x,y))^T.
\end{align}
Here $\mathcal{P}$ and $\mathcal{C}$ are unitary operators and $\mathcal{T}$ is an anti-linear operator \cite{Peskin1995}, \textit{i.e.} $\mathcal{T}(c-\text{number})\mathcal{T}^{-1}=(c-\text{number})^*$. One can check that the mass terms in the Dirac Lagrangian is not invariant under $\mathcal{P}$ or $\mathcal{T}$. However, the combined transformation $\mathcal{PT}$ leaves the mass terms invariant, so $\mathcal{CPT}$ is a symmetry of the Dirac Lagrangian \cite{Deser1982a}. Since the $\gamma^{\mu}$ form three $2\times 2$ matrices and no other matrix anticommutes with them, the chiral symmetry cannot be defined for irreducible representations.
\newline
\newline
\textbf{Reducible representation:} For a reducible representation, let us take the four-component spinor $\psi^T=(\psi^K,\psi^{K'})^T$. As it has been pointed out in Refs. \cite{Gomes1991a,Gomes2009a}, because the free Lagrangian uses only three Dirac matrices, parity, charge conjugation and time-reversal transformations can be implemented by more than one operator
\begin{align}
\mathcal{P}\psi(t,\textbf{r})\mathcal{P}^{-1}&=P_j\psi(t,\textbf{r}'),\\\label{Time-reversalWigner}
\mathcal{T}\psi(t,\textbf{r})\mathcal{T}^{-1}&=T_j\psi(-t,\textbf{r}),\\
\mathcal{C}\psi(t,\textbf{r})\mathcal{C}^{-1}&=C_j(\bar{\psi}(t,\textbf{r}))^T,
\end{align}
where
\begin{align}\label{P1P2parity}
P_1&=-i\Gamma^1\Gamma^3, \ \ \ \ \ \ \ P_2=-\Gamma^1\Gamma^5,\\\label{T1T2Time}
T_1&=-\Gamma^2\Gamma^3, \ \ \ \ \ \ \ \ T_2=-i\Gamma^2\Gamma^5,\\\label{C1C2charge}
C_1&=-i\Gamma^0\Gamma^1, \ \ \ \ \ \ \ C_2=-\Gamma^2,
\end{align}
with $\textbf{r}=(x,y)$ and $\textbf{r}'=(-x,y)$.
\par
We present the transformation properties of some bilinears under $\mathcal{P}$, $\mathcal{C}$ and $\mathcal{T}$ transformations in Tab. \ref{bilinears} . Using these properties, one can easily prove that the massive (or massless) Dirac Lagrangian in $2+1$ dimensions for the reducible representation is invariant under $\mathcal{P}$, $\mathcal{C}$ and $\mathcal{T}$, regardless of which transformation is used, \textit{i.e.} the Dirac Lagrangian is invariant under $P_1$ and $P_2$,  $C_1$ and $C_2$, $T_1$ and $T_2$, or even a linear combination of this operator could be used, with some restrictions \cite{Gomes1991a}. In our Lagrangian, , there are bilinears given by $\bar{\psi} \psi$, $\bar{\psi} \Gamma^{\mu}  \psi$ and $\bar{\psi} \Gamma^{\mu} \Gamma^{35} \psi$, which are independent of $j$. Therefore, any of the operators $P_j$, $C_j$ and $T_j$ can be used to implement $\mathcal{P}$, $\mathcal{C}$ and $\mathcal{T}$, respectively. Surprisingly, the transformation properties of $\bar{\psi} \Gamma^{3}   \psi$,  $\bar{\psi} \Gamma^{5}  \psi$, $\bar{\psi} \Gamma^{\mu} \Gamma^{3}  \psi$ and $\bar{\psi} \Gamma^{\mu} \Gamma^{5}  \psi$ depends on $j$. As a result, we find two non-equivalent realizations of parity, charge conjugation and time-reversal, which is unusual\footnote{In fact, there is an infinite number of non-equivalent realizations of $\mathcal{P}$, $\mathcal{C}$ and $\mathcal{T}$ since any linear combination of the two realizations found is an inequivalent realization.}. For example, the terms $\bar{\psi} \Gamma^{\mu} \Gamma^{3}  \psi$ and  $\bar{\psi} \Gamma^{\mu} \Gamma^{5}  \psi$ appear when a non-Abelian $SU(2)$ gauge field is introduced in graphene \cite{Roy2011,Gopalakrishnan2012a}. In what follows we will be interested only in the Lagrangian density (\ref{DiracLagratwovalleys+-}).
\newline
\def\arraystretch{1.2}
\begin{table*}[ht]
\centering
\begin{tabular*}{0.715\textwidth}{|c|c|c|c|}
\hline
 & $P_j$ & $T_j$ & $C_j$ \\ \hline
$\bar{\psi} \psi(t,\textbf{r})$ & $\bar{\psi} \psi(t,\textbf{r}')$ & $\bar{\psi} \psi (-t,\textbf{r})$ & $\bar{\psi} \psi (t,\textbf{r})$   \\
$\bar{\psi} \Gamma^{\mu}  \psi(t,\textbf{r})$ & $\bar{\psi} \tilde{\Gamma}^{\mu} \psi(t,\textbf{r}')$ & $\bar{\psi} \bar{\Gamma}^{\mu} \psi (-t,\textbf{r})$ & $-\bar{\psi} \Gamma^{\mu} \psi (t,\textbf{r})$   \\
$\bar{\psi} \Gamma^{\mu} \Gamma^{35} \psi(t,\textbf{r})$ & $-\bar{\psi} \tilde{\Gamma}^{\mu} \Gamma^{35} \psi(t,\textbf{r}')$ & $-\bar{\psi} \bar{\Gamma}^{\mu} \Gamma^{35} \psi (-t,\textbf{r})$ & $\bar{\psi} \Gamma^{\mu} \Gamma^{35} \psi (t,\textbf{r})$ \\
$\bar{\psi} i \Gamma^{35} \psi(t,\textbf{r})$ & $-\bar{\psi} i \Gamma^{35} \psi(t,\textbf{r}')$ & $-\bar{\psi} i \Gamma^{35} \psi (-t,\textbf{r})$ & $\bar{\psi} i \Gamma^{35} \psi (t,\textbf{r})$ \\
$\bar{\psi} \Gamma^{3}  \psi(t,\textbf{r})$ & $(-1)^j\bar{\psi} \Gamma^{3} \psi(t,\textbf{r}')$ & $(-1)^{j+1}\bar{\psi} \Gamma^{3}\psi (-t,\textbf{r})$ & $(-1)^{j+1}\bar{\psi} \Gamma^{3} \psi (t,\textbf{r})$   \\
$\bar{\psi} \Gamma^{5}  \psi(t,\textbf{r})$ & $(-1)^{j+1}\bar{\psi} \Gamma^{5} \psi(t,\textbf{r}')$ & $(-1)^{j}\bar{\psi} \Gamma^{5}\psi (-t,\textbf{r})$ & $(-1)^{j}\bar{\psi} \Gamma^{5} \psi (t,\textbf{r})$   \\
$\bar{\psi} \Gamma^{\mu} \Gamma^{3}  \psi(t,\textbf{r})$ & $(-1)^{j}\bar{\psi} \tilde{\Gamma}^{\mu} \Gamma^{3}  \psi(t,\textbf{r}')$ & $(-1)^{j+1}\bar{\psi}\bar{\Gamma}^{\mu} \Gamma^{3}  \psi (-t,\textbf{r})$ & $(-1)^{j+1}\bar{\psi} \Gamma^{\mu}\Gamma^{3}  \psi (t,\textbf{r})$   \\
$\bar{\psi} \Gamma^{\mu} \Gamma^{5}  \psi(t,\textbf{r})$ & $(-1)^{j+1}\bar{\psi} \tilde{\Gamma}^{\mu} \Gamma^{3}  \psi(t,\textbf{r}')$ & $(-1)^{j}\bar{\psi} \bar{\Gamma}^{\mu} \Gamma^{3}  \psi (-t,\textbf{r})$ & $(-1)^{j}\bar{\psi} \Gamma^{\mu}\Gamma^{3}  \psi (t,\textbf{r})$
  \\ \hline
\end{tabular*}
\caption{$\mathcal{P}$, $\mathcal{C}$ and $\mathcal{T}$ transformation properties of some bilinears, here
$\tilde{\Gamma}^{\mu}=\{\Gamma^{0},-\Gamma^{1} ,\Gamma^{2}\}$ and $\bar{\Gamma}^{\mu}=\{\Gamma^{0},-\Gamma^{i}\}$.}\label{bilinears}
\end{table*}
\par
It should be noted that in the literature one can find two different transformations which are defined as time-reversal. One, $\hat{T}$, which acts as $\hat{\mathcal{T}}\psi(t,x,y)\hat{\mathcal{T}}^{-1}=\hat{T}\psi^{*}(-t,\textbf{r})$. The other, $T$, which is the one considered here, Eq. (\ref{Time-reversalWigner}), and is referred as Wigner time-reversal. The latter transformation was defined consistently with what has been done in four dimensions by Weinberg (Ch. 5. in \cite{Weinberg1996}), and Peskin and Schroeder (Ch. 3. in \cite{Peskin1995}). Moreover, the one that concerns the $\mathcal{CPT}$ theorem is $T$ (for a detailed discussion, see sec. 11.6. in \cite{Schwartz2014}).
\par
The transformation properties of the electromagnetic potential are \cite{Deser1982a}
\begin{align}
\mathcal{P}A_0(t,x,y)\mathcal{P}^{-1}&=A_0(t,-x,y),\nonumber\\
\mathcal{P}A_1(t,x,y)\mathcal{P}^{-1}&=-A_1(t,-x,y),\nonumber\\
\mathcal{P}A_2(t,x,y)\mathcal{P}^{-1}&=A_2(t,-x,y),\nonumber\\
\mathcal{T}A_0(t,x,y)\mathcal{T}^{-1}&=A_0(-t,x,y),\nonumber\\
\mathcal{T}\vec{A}(t,x,y)\mathcal{T}^{-1}&=-\vec{A}(-t,x,y),\nonumber\\
\mathcal{C}A_{\mu}(t,x,y)\mathcal{C}^{-1}&=-A_{\mu}(t,x,y),
\end{align}
which leave the Lagrangian invariant. For pseudo-magnetic potential we should have
\begin{align}
\mathcal{P}a_1^{35}(t,x,y)\mathcal{P}^{-1}&=a_1^{35}(t,-x,y),\nonumber\\
\mathcal{P}a_2^{35}(t,x,y)\mathcal{P}^{-1}&=-a_2^{35}(t,-x,y),\nonumber\\
\mathcal{T}\vec{a}^{35}(t,x,y)\mathcal{T}^{-1}&=\vec{a}^{35}(-t,x,y),\nonumber\\
\mathcal{C}\vec{a}^{35}(t,x,y)\mathcal{C}^{-1}&=\vec{a}^{35}(t,x,y).
\end{align}
so that the interaction $\bar{\psi}\Gamma^i\Gamma^{35}a_i^{35}\psi$ in the Lagrangian is invariant.
\par
For reducible representations, the transformation $\psi\rightarrow e^{i\alpha_{\mu}\tilde{\sigma}^{\mu}}\psi$ leaves invariant the kinetic term, where
the generators $\tilde{\sigma}^{\mu}=\sigma^{\mu}\otimes I_{2\times2}=\{I_{4\times 4}, -i\Gamma^3,-\Gamma^5,-i\Gamma^{35}\}$ are the generators of a global $U(2)$ symmetry, with $\sigma^0\equiv I_{2\times 2}$ and the $\alpha_{\mu}$ are taken as constants. The mass term $m\bar{\psi}\psi$ breaks this global symmetry down to $U(1)\times U(1)$ symmetry, whose generator are $I_{4\times 4}$ and $-i\Gamma^{35}$ \cite{Das1997}. However, when the mass vanishes, the quantum corrections generate a vacuum expectation value of $\bar{\psi}\psi$ (to be precise $[\bar{\psi},\psi]/2$, see below), then, the symmetry would have broken down to $U(1)\times U(1)$.
\par
It should be noted that besides the usual mass term $m\bar{\psi}\psi$, there is a mass term $m_{\tau}\bar{\psi}\frac{[\Gamma^3,\Gamma^5]}{2}\psi=m_{\tau}\bar{\psi}i\Gamma^{35}\psi$ known as Haldane mass term \cite{Haldane1988a}, which is invariant under the $U(2)$ symmetry. However, this term breaks parity and time-reversal symmetries (see  Tab. \ref{bilinears}).
\section{Furry picture}\label{secFurry}
In this section we present the Furry picture based on the quantized solutions of the Dirac equation, and we generalize what has been done in Refs. \cite{Dunne1996,Das1996} for a Dirac equation in the presence of a real magnetic field to the case in which the Dirac equation is in the presence of real and pseudomagnetic fields. In static background gauge fields, the Dirac equation (\ref{DiracEpm}) can be rewritten as
\begin{equation}
\left( \begin{array}{cc}
E\mp m & -(D_1^{\pm}-iD_2^{\pm})\\
(D_1^{\pm}+iD_2^{\pm}) & E\pm m \\
\end{array} \right)\psi=0.
\end{equation}
There are two possible solutions depending on the threshold states ($|E|=\pm m$). The positive-energy solutions ($\psi^{(+)}$) are
\begin{eqnarray}\nn
\psi^{(+)}_{\pm,1}&=&e^{-i|E|t}\sqrt{\frac{|E|\pm m}{2|E|}}\left( \begin{array}{c}
f\\
-\frac{D_1^{\pm}+iD_2^{\pm}}{|E|\pm m}f\\
\end{array} \right), \ \ \ \text{or} \\ \label{PositiveMGfg}
\psi^{(+)}_{\pm,2}&=&e^{-i|E|t}\sqrt{\frac{|E|\mp m}{2|E|}}\left( \begin{array}{c}
\frac{D_1^{\pm}-iD_2^{\pm}}{|E|\mp m}g\\
g\\
\end{array} \right),
\end{eqnarray}
where $\psi^{(+)}_{+,i}$ refers to the positive-energy solution in the representation $\mathcal{R}_1$ and $\psi^{(+)}_{-,i}$ refers to the positive-energy solution in the representation $\mathcal{R}_2$. The negative-energy solutions ($\psi^{(-)}$) are
\begin{eqnarray}\nn
\psi^{(-)}_{\pm,1}&=&e^{+i|E|t}\sqrt{\frac{|E|\mp m}{2|E|}}\left( \begin{array}{c}
f\\
-\frac{D_1^{\pm}+iD_2^{\pm}}{|E|\mp m}f\\
\end{array} \right), \ \ \ \text{or} \\ \label{NegativeMGfg}
\psi^{(-)}_{\pm,2}&=&e^{+i|E|t}\sqrt{\frac{|E|\pm m}{2|E|}}\left( \begin{array}{c}
\frac{D_1^{\pm}-iD_2^{\pm}}{|E|\pm m}g\\
g\\
\end{array} \right),
\end{eqnarray}
where $f$ and $g$ are two functions such that
\begin{align}
-(D_1^{\pm}-iD_2^{\pm})(D_1^{\pm}+iD_2^{\pm})f&=(E^2-m^2)f,\\
-(D_1^{\pm}+iD_2^{\pm})(D_1^{\pm}-iD_2^{\pm})g&=(E^2-m^2)g.
\end{align}
Note that the threshold states $|E|=m$ and $|E|=-m$ must be specified separately. When $|E|=m$ is a positive (negative) energy solution, the negative (positive) energy threshold is excluded, because of the factor $1/\sqrt{|E|-m}$ \cite{Dunne1996}. For example, for the valley $\vec{K}$, or equivalently the representation $\mathcal{R}_1$, the positive-energy solutions for $|E|=m>0$ and $|E|=m<0$ are respectively
\begin{equation}\label{PositiveMGfg0}
\psi^{(+0)}_{+,1}=e^{-i|m|t}\left( \begin{array}{c}
f^{(0)}\\
0\\
\end{array} \right), \ \ \ \ \ \psi^{(+0)}_{+,2}=e^{-i|m|t}\left( \begin{array}{c}
0\\
g^{(0)}\\
\end{array} \right),
\end{equation}
where $f^{(0)}(x,y)$ satisfies the first-order threshold equation
\begin{equation}\label{thresholdequf}
(D_1^{+}+iD_2^{+})f^{(0)}=0,
\end{equation}
and $g^{(0)}(x,y)$ satisfies
\begin{equation}\label{thresholdequg}
(D_1^{+}-iD_2^{+})g^{(0)}=0.
\end{equation}
It turns out that if the solutions of (\ref{thresholdequf}) are normalizable, then the solutions of (\ref{thresholdequg}) are not, and vice versa \cite{Dunne1996,Aharonov1979a}. Now, in the absence of pseudo-magnetic field ($a^{35}_i=0$), one has that $D=D^+=D^-$. Thus, if $\psi^{(+0)}_{+,1}$ $\left(\psi^{(+0)}_{+,2}\right)$ is a positive-energy solution for the valley $\vec{K}$, then the valley $\vec{K}'$ only has the negative-energy solution $\psi^{(-0)}_{-,2}$ $\left(\psi^{(-0)}_{-,1}\right)$. This leads to the well-known asymmetry in the spectrum of the states. Remarkably, this does not necessarily happen when there is a pseudo-magnetic field, since $D^+\neq D^-$. Additionally, if the solutions of (\ref{thresholdequf}) are normalizable, this does not imply that the solutions of $(D_1^{+}-iD_2^{+})g^{(0)}=0$ are not. Therefore, both valleys may have positive (or negative) energy states simultaneously. The Appendix (\ref{ap:ConsRPMF}) illustrates this point in the case of constant real and pseudomagnetic fields.
\par
One can calculate the vacuum condensate $\langle \bar{\psi}\psi \rangle$ (pairing between fermions and antifermions in the vacuum)
in $2+1$ dimensions by expanding out the fermion field in a complete orthonormal set of the positive- and negative-energy solutions ($i=\mathcal{R}_1,\mathcal{R}_2$)
\be\label{psicampo}
\Psi_i(\vec{x},t)=\SumInt_{n}
\SumInt_p
[a_{i,n,p}\psi^{(+)}_{i,n,p}+b^{\dag}_{i,n,p}\psi^{(-)}_{i,n,p}].
\ee
The solutions are labeled by two quantum numbers $(n,p)$, in which the label $n$ refers to the eigenvalue $E_n$, whilst the label $p$ distinguishes between degenerate states. In general, both $n$ and $p$ may take discrete and/or continuous values \cite{Dunne1996}. The $a_{i,n,p}$ and $b^{\dag}_{i,n,p}$ are the fermion annihilation operator and antifermion creation operator, respectively, which obey the anticommutation relations
	\be\label{conmutationrela}
\{a_{i,n,p},a^{\dag}_{j,n',p'}\}=\{
b_{i,n,p},b^{\dag}_{j,n',p'}\}=\delta_{ij}\bar{\delta}_{nn'}\bar{\delta}_{pp'},.
\ee
where $\bar{\delta}_{\alpha,\alpha'}$, is the Kronecker delta if $\alpha$ takes discrete values, or is the Dirac delta if it takes continuous values. Using the commutation relations (\ref{conmutationrela}), the vacuum expectation value $\langle \bar{\Psi}_i\Psi_i \rangle$ can be written as
\be\nn
\langle 0| \bar{\Psi}_i(x)\Psi_i(x)|0 \rangle \equiv \text{tr} \langle \bar{\Psi}_{i,\alpha}\Psi_{i,\beta} \rangle =\SumInt_{n}
\SumInt_p \bar{\psi}^{(-)}_{i,n,p}(x) \psi^{(-)}_{i,n,p}(x),\\\label{psiAPsiAG}
\ee
\textit{i.e} the fermion condensate is a sum over occupied negative-energy states. The $\text{tr}$ is over the spinorial indices $\{\alpha,\beta\}$. Let us also write the condensate $\langle \Psi_i \bar{\Psi}_i \rangle$, which will be relevant to what follows, as
\be\nn
\langle 0| \Psi_i(x)\bar{\Psi}_i(x)|0 \rangle \equiv \text{tr} \langle \Psi_{i,\beta} \bar{\Psi}_{i,\alpha} \rangle=\SumInt_{n}
\SumInt_p \bar{\psi}^{(+)}_{i,n,p}(x) \psi^{(+)}_{i,n,p}(x),\\ \label{PsiAGpsiA}
\ee
\textit{i.e} this condensate is a sum over occupied positive-energy states.
\par
In the \ref{ap:MagConApen}, we compare the calculation of fermionic condensate in the Furry picture via fermion propagator. We prove that the trace of the fermion propagator evaluated at equal space-time points must be understood as the expectation value of the commutator of two field operators. Thus, the Schwinger's choice \cite{Schwinger1951}, which is equivalent to perform the coincidence limit symmetrically on the time coordinate \cite{Dittrich1985}, \textit{i.e.}
\be\nn
\frac{\langle 0|[\bar{\Psi}_i,\Psi_i]|0 \rangle}{2}&=&-\text{tr} S_F(x,x)\\\label{OrderConmutator}
&\equiv &
-\left(\text{tr} \lim_{\substack{x_0\rightarrow y_0^{+}\\ \mathbf{x}\rightarrow
\mathbf{y}}} S_F(x,y)-\text{tr} \lim_{\substack{x_0\rightarrow y_0^{-}\\ \mathbf{x}\rightarrow \mathbf{y}}} S_F(x,y)\right).
\ee
Besides, we argue that this must be the order parameter for chiral (or parity) symmetry breaking and not the fermionic condensate, as is commonly assumed in the literature.
\section{Fermion condensate}\label{secCondesateM}
Chiral symmetry breaking in ($2+1$)- and
($3+1$)-dimensional theories has been a subject of intense
scrutiny over the past two decades
\cite{Gusynin1994,Gusynin1995a,Gusynin1995b,Gusynin1995c,
Gusynin1996a,Gusynin1999d,Dunne1996,Dittrich1997,Cea1997,Cea1998,
Farakos1998a,Farakos1998b,Lasinio1999,Cea2000,Anguiano2005,Raya2010,
Cea2012a,Boyda2014,Ayala2006,Ferrer2010,Ayala2010,Khalilov2015}.
In the presence of a uniform magnetic field, the appearance of a
nonvanishing chiral condensate $\langle \bar{\Psi}\Psi \rangle
\neq 0$ in the limit $m\rightarrow0$, produce spontaneous chiral
symmetry breaking \cite{Gusynin1994,Gusynin1995a,Gusynin1995b}.
For example \cite{Gusynin1994,Gusynin1995c}, in the Nambu-Jona-Lasinio (NJL) 
model, the spontaneous symmetry breaking occurs when the coupling constant exceeds some critical value,
\textit{i.e.} when $\lambda>\lambda_c$. With an external uniform
magnetic field $\lambda_c\rightarrow0$, independent of the
intensity of the magnetic field $B$, the magnetic field is a
strong \textit{catalyst} of chiral symmetry breaking (see Ref.
\cite{Shovkovy2013} for review).
\par
The exact expression for a fermion propagator in an external
magnetic field in $3+1$ dimensions was found for the first time by
Schwinger using the proper-time formalism \cite{Schwinger1951}. For $2+1$ dimensions, the fermion 
propagator was presented in the momentum representation by Gusynin
\textit{et al.} in \cite{Gusynin1994}. In \cite{Gusynin1994,Gusynin1995a,Gusynin1996a}, the vacuum condensate
was computed in the reducible representation (fourcomponent spinor) using the expression of the fermion propagator in the presence of an uniform magnetic field. On the other hand, Das and Hott introduced an alternative derivation of the magnetic condensate using the Furry Picture. This method has been used to calculate the magnetic vacuum condensate in $3+1$ dimensions \cite{Anguiano2007}, at finite temperature \cite{Das1996,Das1997,Cea1998,Cea2000}, and as well as in the presence of parity-violating mass terms \cite{Anguiano2007}. In appendix \ref{ap:MagConFurryP}, we compute the vacuum condensate in the presence of an external magnetic field for the two irreducible representations and show that these two methods are consistent if we take the definition of the propagator in equal times, as the one introduced by Schwinger in \cite{Schwinger1951}. Furthermore, we discuss why the vacuum expectation value of the commutator of two field operators is the appropriate order parameter to describe chiral (or parity) symmetry breaking.
\par
In order to study the effect of strains, in the following we consider a sample of graphene in the presence of constant real ($B$) and pseudo ($b$) magnetic fields\footnote{We noted that the pseudomagnetic field study here is mathematically equivalent to considering a Dirac oscillator potential in $(2+1)$-dimensions  \cite{Quimbay2013}. Consequently, a constant pseudomagnetic field can be seen as a physical realization of the two-dimensional Dirac oscillator.}. In this case, we choose $\vec{A}=(0,Bx)$ and $\vec{a}^{35}=(0,bx)$. The explicit solution can be found in the (\ref{ap:ConsRPMF}). In a similar way shown in \ref{ap:MagConFurryP}, we compute the vacuum expectation value of the commutator in the two irreducible representations for arbitrary values of $m$, $eB$ and $eb$, i.e.,
\begin{align}\nn
&\frac{1}{2}\langle[\bar{\Psi}_{\pm},\Psi_{\pm}] \rangle_{B}
=-\text{sgn}(m)\frac{|eB\pm eb|}{4\pi}-\frac{|eB\pm eb|}{2\pi}\sum_{n=1}^{\infty} \frac{m}{|E^{\pm}_n|}\\\label{BPsiPsiB}
&=\text{sgn}(m)\frac{|eB\pm eb|}{4\pi}-m\frac{\sqrt{2|eB\pm eb|}}{4\pi}\zeta\left(\frac{1}{2},\frac{m^2}{2|eB\pm eb|}\right),
\end{align}
with $|E_n^{\pm}|=\sqrt{m^2+2|eB\pm eb|n}$, here the $(-)$ sign refers to the representation $\mathcal{R}_1$ (valley $\vec{K}$), whereas $(+)$ refers to the representation $\mathcal{R}_2$ (valley $\vec{K}'$). $\zeta(s,q)$ is the Hurwitz zeta function defined by
\be
\zeta(s,q)=\sum_{n=0}^{\infty}\frac{1}{(n+q)^{s}}.
\ee
The commutator can be rewritten in an integral representation as
\begin{eqnarray}\label{CondensataZeroDiv1}\nn
&&\frac{1}{2}\langle[\bar{\Psi}_{\pm},\Psi_{\pm}] \rangle_{B,b}\\
&&=-\frac{m}{4 \pi^{\frac{3}{2}}}\int_0^{\infty}dt e^{-m^2 t}t^{-\frac{1}{2}}|eB\pm eb|\coth (|eB\pm eb|t),
\end{eqnarray}
where we have used that
\begin{eqnarray}\nn
&&\int_0^{\infty}dt e^{-m^2 t}t^{-\frac{1}{2}}|\omega|\coth (|\omega|t)\\\label{integralrepre}
&&=(2|w|\pi)^{\frac{1}{2}}\left(\zeta\left(\frac{1}{2},\frac{m^2}{2|w|}\right)-\frac{|w|^{\frac{1}{2}}}{2^{\frac{1}{2}}|m|}\right),
\end{eqnarray}
which can be obtained after regularization with the $\epsilon-$inte\-gration technique \cite{Dittrich2000}.
Although Eq. (\ref{CondensataZeroDiv1}) is divergent, the divergences are already present for zero external field
\begin{align}\label{CondensataZeroDiv}
\frac{1}{2}\langle[\bar{\Psi}_{\pm},\Psi_{\pm}] \rangle_{0,0}
=-\frac{m}{4 \pi^{\frac{3}{2}}}\int_0^{\infty}dt e^{-m^2 t}t^{-\frac{3}{2}}.
\end{align}
Therefore, by subtracting out the vacuum part, a finite result is obtained  \cite{Dittrich2000,Ayala2010,Farakos1998b}
\begin{align}\nn
\mu^{\pm}&=\frac{1}{2}\langle[\bar{\Psi}_{\pm},\Psi_{\pm}] \rangle_{B,b}-\frac{1}{2}\langle[\bar{\Psi}_{\pm},\Psi_{\pm}] \rangle_{0,0}\\\label{integralcconde}
&
=-\frac{m}{4 \pi^{\frac{3}{2}}}\int_0^{\infty}dt e^{-m^2 t}t^{-\frac{1}{2}}\left(|eB\pm eb|\coth (|eB\pm eb|t)-\frac{1}{t}\right).
\end{align}
\begin{figure} 
\begin{minipage}{\columnwidth}
\centering
\includegraphics[width=8.1cm]{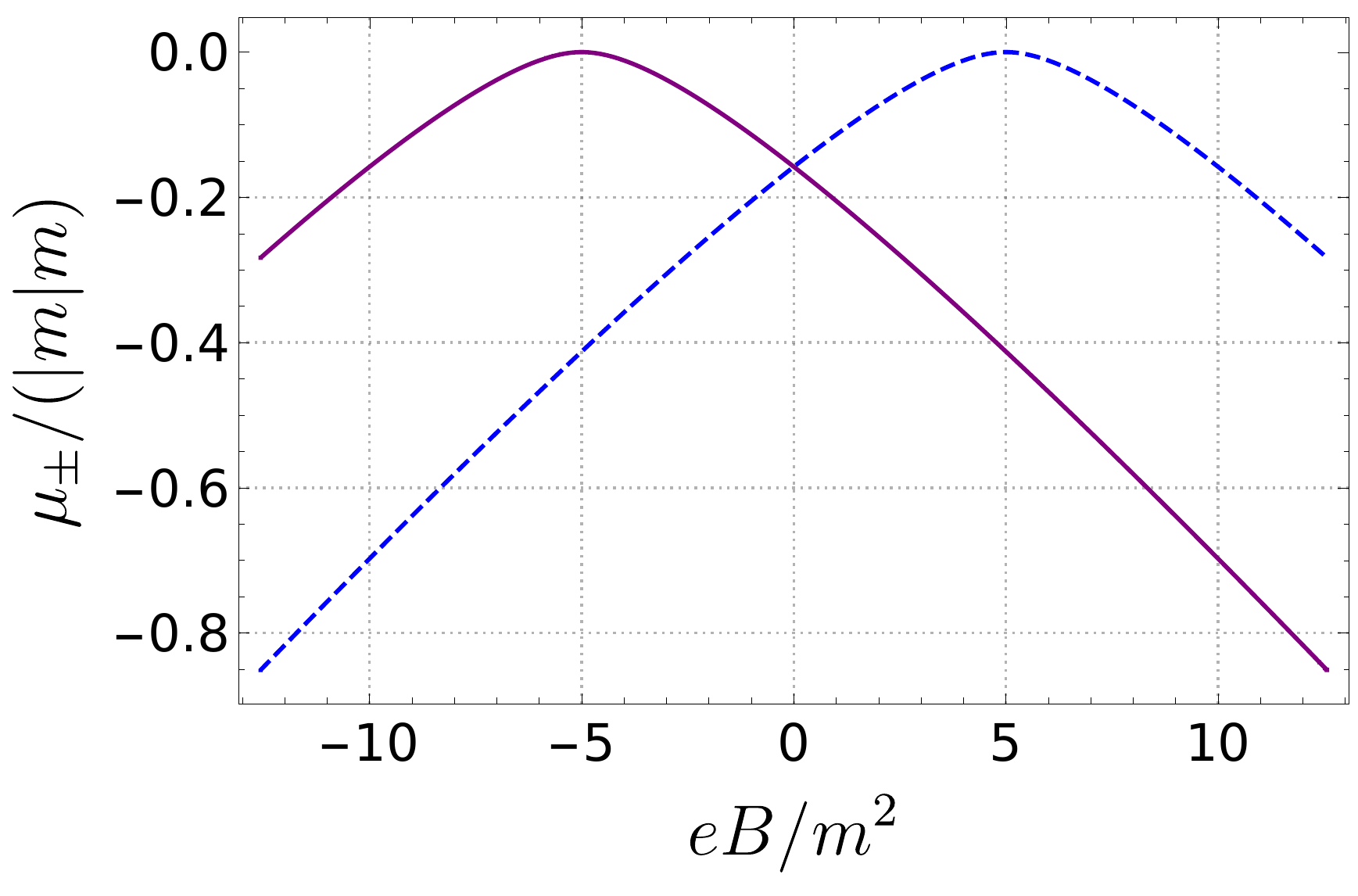}
\end{minipage}
\caption{(color online). The $c-$condensates as a function of a external magnetic field, $\mu_-/(m|m|)$ (dashed line) and $\mu_+/(m|m|)$ (continuous line), for $eb/m^2=5$. Since $\mu^{\pm}(eB\pm eb,m)/(m|m|)=\mu^{\pm}\left(\frac{eB\pm eb}{m^2},1\right)$ it is satisfied, when the mass is changed only a widening of the ``parabolic'' form of the function occurs.}
\label{ccondensates}
\end{figure}
To evaluate in a simple form Eq. (\ref{CondensataZeroDiv}), first note that although this integral is divergent, Eq. (\ref{BPsiPsiB}) has a finite limit as $|eB\pm eb|\rightarrow 0$ if we used the analytic continuation of the Hurwitz zeta function so $\frac{1}{2}\langle[\bar{\Psi}_{\pm},\Psi_{\pm}] \rangle_{0,0}=\text{sgn}(m)\frac{m}{2 \pi}$, which coincides with the regularization using the $\epsilon$-integration technique \cite{Dittrich2000}. Henceforth, we will refer to $\mu^{\pm}$ as the $c-$ condensates, which in terms of the analytically continue Hurwitz zeta function are given by\footnote{For numerical calculations, 
instead of utilizing the integral (\ref{integralcconde}), the $c-$condensates can be evaluated in a much more efficient way using the Hurwitz zeta functions.}
\begin{eqnarray}\nn
\mu^{\pm}
&=&-\text{sgn}(m)\frac{|eB\pm eb|}{4\pi}\\
&&-m\frac{\sqrt{2|eB\pm eb|}}{4\pi}\zeta\left(\frac{1}{2},1+\frac{m^2}{2|eB\pm eb|}\right)-|m|\frac{m}{2\pi}.
\end{eqnarray}
One can show that there is no critical value of the fields in which the $\mu^{\pm}/m$ changes sign. Notably, if $B\neq0$ or $b\neq 0$, the values of the $\mu^{\pm}$ for each valley are different and when $|B|=|b|$ one of the two is zero, as showed in Fig. \ref{ccondensates}. Finally, let us take the limit $m\rightarrow 0$
\be\label{ccondenm02x2}
\mu^{\pm}=-\text{sgn}(m)\frac{|eB\pm eb|}{4\pi}.
\ee
As we will see below, this is related to a breaking of chiral, parity and time-reversal symmetries.

\subsection{Dynamical mass}
\begin{figure} 
\begin{minipage}{\columnwidth}
\centering
\includegraphics[width=8.1cm]{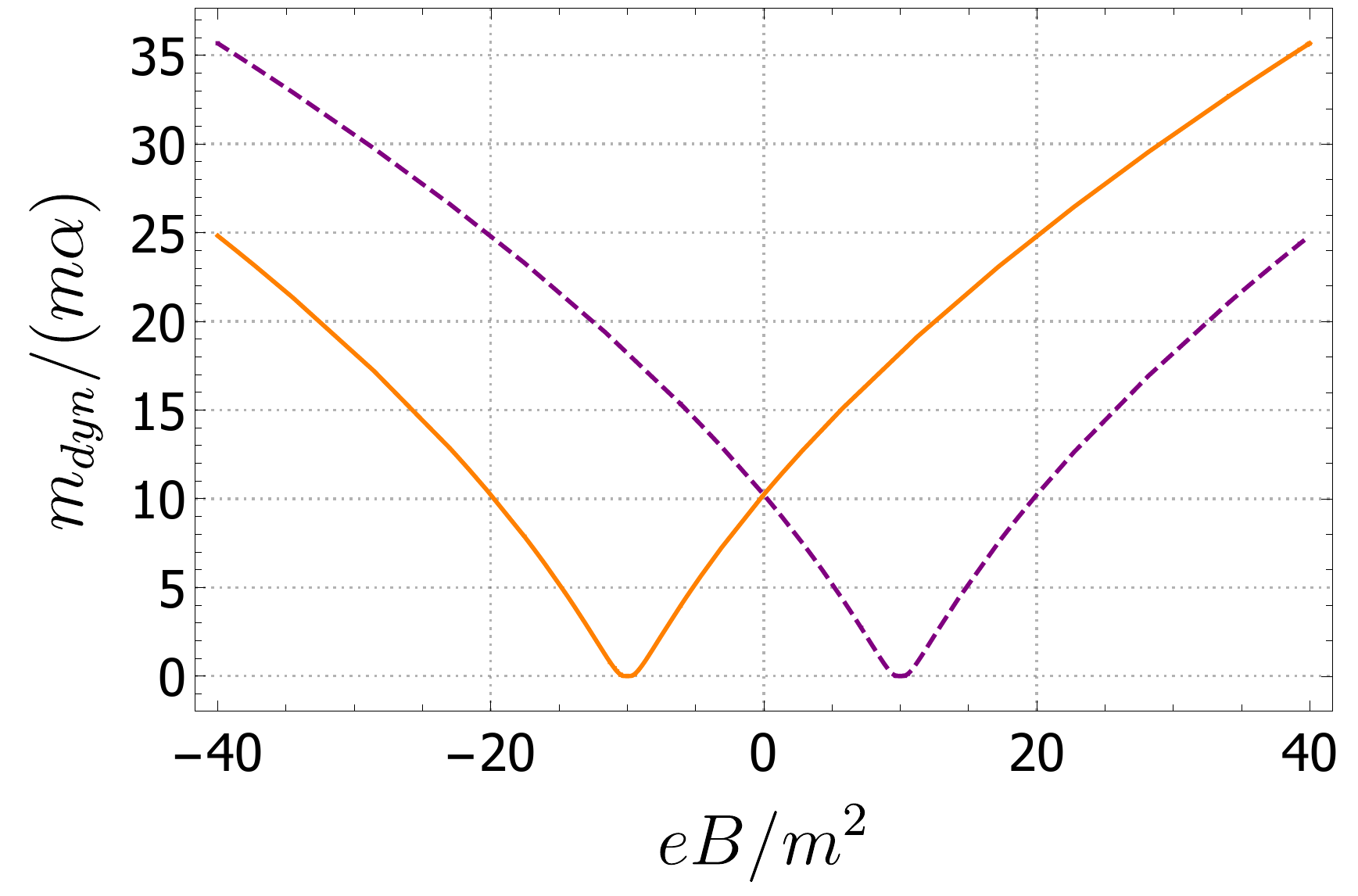}
\end{minipage}
\caption{(color online) Dynamical masses in the constant-mass approximation versus external magnetic field, $m^{+}_{dyn}/(m\alpha)$ (continuous line) and $m^{-}_{dyn}/(m\alpha)$ (dashed line), for $eb/m^2=10$.}
\label{dinamicalmasses}
\end{figure}
Dynamical mass generation in QED$_{2+1}$ has been a subject of study in the past three decades \cite{Hoshino1989a,Gusynin1994,Shpagin1996,Farakos1998a,Farakos1998b,Raya2010,khalilov2019a}. As shown in Ref. \cite{Raya2010}, the dynamical mass with a two-component fermion in a uniform magnetic field, in the so-called constant-mass approximation, is
\begin{align}
m_{dyn}^{2+1}=2\alpha W\left(\frac{e^{-\gamma_E/2}\sqrt{2|eB|}}{2\alpha}\right),
\end{align}
with $\gamma_E$ the Euler constant, $\alpha=e^2/(4\pi)$ and $W(x)$ the Lambert $W$ function. In the latter formula, it is necessary that $|eB|\gg m_{dyn}^2$ for consistency.
For weak magnetic fields, the dynamical mass has a quadratic behavior in the magnetic field, $m_{dyn}^{2+1}=m_0+m_2(|eB|)^2+\dots$ \cite{Farakos2000a}. Futhermore, the radiative corrections to the
mass of a charged fermion when it occupies the lowest Landau level in RQED$_{4,3}$ was recently computed in the one-loop approximation in Refs. \cite{khalilov2019a,Machet2018a}. Its associated equation reads
\begin{eqnarray}\nn
m_{dyn}^{RQED}=\frac{e^{2}}{4 \pi} \sqrt{|e B|}\left[\sqrt{2} \pi^{3 / 2} \operatorname{erfc}\left(\frac{1}{\sqrt{l}}\right)-\frac{10}{3 \sqrt{l}} \Gamma\left(0, \frac{1}{l}\right)\right],\\
\end{eqnarray}
with $l=\frac{|eB|}{m^2}$ and where erfc$(z)$ and $\Gamma(z)$ are the complementary error function and upper incomplete gamma function, respectively. Significantly, the dynamical mass does not vanish even at the limit of zero bare mass ($m\rightarrow 0$) \cite{Machet2018a,khalilov2019a}
\begin{equation}
m_{dyn}^{RQED}=\frac{e^{2}}{2 \sqrt{2}} \sqrt{\pi|e B|}.
\end{equation}
In graphene, while the photon propagates in $3$ spatial dimensions, the fermions are localized on $2$ spatial dimensions, because of this, RQED$_{4,3}$ is an appropriate model to describe the low-energy physics for this system. Nevertheless, as mentioned above, the coupling constant is large, hence, this perturbative result is not necessarily accurate \cite{Kolomeisk2015a,khalilov2019a}. In the following, because of the lack of a better estimate, we use this result to determine the dynamical mass in graphene.
\par
In general, the dynamical mass should be given by $m_{dyn}=g(|eB|)$, where $g$ is a general function. It is straightforward to extend this result to include the pseudomagnetic field. Hence, the dynamical mass would read as $m_{dyn}=g(|eB\pm eb|)$ and we thus obtain
\begin{eqnarray}\nn
m_{dyn}^{\pm}&=&\alpha \sqrt{|e B\pm eb|} \left[\sqrt{2} \pi^{3 / 2} \operatorname{erfc}\left(\frac{m}{\sqrt{|e B\pm eb|}}\right)\right.\\
&&\left.-\frac{10m}{3 \sqrt{|e B\pm eb|}} \Gamma\left(0, \frac{m^2}{|e B\pm eb|}\right)\right],
\end{eqnarray}
with $\alpha=e^2/(4\pi)$.
According to these findings, the dynamical fermion mass is different for each valley, $m^{+}_{dyn}$ for $\vec{K}'$ and $m^{-}_{dyn}$ for $\vec{K}$ (see Fig. \ref{dinamicalmasses}). This is not surprising since the $c-$condensates, $\mu^{+}$ and $\mu^{-}$, are different if a pseudomagnetic field is included. Furthermore, it is possible to construct a Lagrangian that describes two species of fermions, each with different mass, introducing the usual mass term ($m$) and a Haldane mass term ($m_{\tau}$)
\begin{align}
\mathcal{L}_m=m\bar{\psi}\psi+m_{\tau}\bar{\psi}i\Gamma^{35}\psi.
\end{align}
In this case, the two masses will be $m\pm m_{\tau}$ and $\psi$ is taken as a four-component spinor. This result implies that the interplay between the real and pseudomagnetic fields allows us to dynamically generate these two terms. With the help of Eq. (\ref{ccondenm02x2}), one can realize that
\begin{eqnarray}\nn
&&\frac{1}{2}\langle[\bar{\psi},\psi] \rangle_{B,b}-\frac{1}{2}\langle[\bar{\psi},\psi] \rangle_{0,0}\\
&&=-\frac{\text{sgn}(m)}{4\pi}(|eB+eb|+|eB-eb|),
\end{eqnarray}
while in the limit $m\rightarrow 0$, we have
\begin{eqnarray}\nn
&&\frac{1}{2}\langle[\bar{\psi},i\Gamma^{35}\psi] \rangle_{B,b}-\frac{1}{2}\langle[\bar{\psi},i\Gamma^{35}\psi] \rangle_{0,0}\\
&&=-\frac{\text{sgn}(m)}{4\pi}(|eB+eb|-|eB-eb|).
\end{eqnarray}
Therefore, the usual mass term is always generated independently of $B$ and $b$, whereas the Haldane mass term is only generated if $B\neq 0$ and $b\neq 0$ simultaneously. Note that when $B=b$ (or $B=-b$), one of the $c-$condensates is zero and the dynamical mass of this valley will be independent of $B$ and $b$.
\par
Finally, it is important to realize that for zero pseudomagnetic field, the mass term in irreducible representations breaks parity and time-reversal symmetries, while in a reducible representation chirality is broken. Thus, for irreducible representations, the $c-$ condensate ($\mu=\mu^{+}=\mu^{-}$) is the order parameter of the dynamical parity and there is a time-reversal symmetry breaking. In contrast, dynamical symmetry breaking occurs in reducible representations. In reducible representations, however, non-zero magnetic and pseudo magnetic fields produce a dynamical symmetry breaking, not only of the chiral symmetry but also of the parity and time-reversal symmetries\footnote{In Ref. \cite{Herbu2008a}, it had been suggested that the flux of the non-Abelian pseudomagnetic field catalyzes the time-reversal symmetry breaking.}. The reason for this is that by including pseudomagnetic field, the dynamical mass is different for each valley
since a Haldane mass term (which breaks parity and time reversal) is generated\footnote{The Haldane mass term, for example, can also be dynamically generated in graphene at sufficiently large strength of the long-range Coulomb interaction \cite{Gonzalez2013}.}.
\section{One-loop effective action and magnetization}\label{oneloopeamagnetization}
In this section we compute the effective action and the magnetization
in the presence of uniform real and pseudomagnetic fields. We consider the fermionic
part of the generating functional for each valley
\begin{eqnarray}
&&Z_{\pm}=e^{iW_{\pm}}\\\nn
&&=\int  \mathcal{D} \bar{\Psi}_{\pm} \mathcal{D}\Psi_{\pm}
\exp \left( i\int d^3x  [ \bar{\Psi}_{\pm} (i \slashed{\partial}
+e\slashed{A}\mp e\slashed{a}^{35}-m ) \Psi_{\pm}]	\right).
\end{eqnarray}
Then, we introduce the one-loop effective Lagrangian $\mathcal{L}^{(1)}_{\pm}$ via
$\text{ln} Z_{\pm} = i \int d^3x \mathcal{L}^{(1)}_{\pm} (x)$. In the presence of a static background field, the one-loop effective action is proportional to the vacuum energy (Ch. 16.  in \cite{Weinberg1996}). The vacuum energy of the Dirac energy field can be computed using the formula \cite{Gunter1986}
\begin{equation}\label{EnerVaccumForm}
E_{\text{vac}}=\frac{1}{2}\left(-\sum_{E_n>0}E_n+\sum_{E_n<0}E_n\right),
\end{equation}
which depends upon the zero-point energies of both positive- and negative-energy states\footnote{Provided that for each eigenvalue $E_n$, there is an eigenvalue $-E_n$, then the two sums in Eq. (\ref{EnerVaccumForm}) reduces to the sum over the Dirac sea
$$E_{\text{vac}}=\sum_{E_n<0}E_n.$$
This equation is always satisfied by a charge conjugation invariant background; however, this is not our case. The use of this equation in a magnetic field background has led to erroneous conclusions in Ref. \cite{Cea1985,Cea2000}.}.
In our case, it is straightforward to obtain the density vacuum energy for each valley
\begin{eqnarray}\nn
&&\mathcal{E}^{\pm}_{\text{vac}}(B,b)=\frac{E^{\pm}_{\text{vac}}}{A}\\
&&=-\frac{|eB\pm eb|}{4 \pi}|m|-\frac{|eB\pm eb|}{2 \pi}\sum_{n=1}^{\infty}\sqrt{m^2+2|eB\pm eb|n}\\\nn
&&=\frac{|eB\pm eb|}{4 \pi}|m|-\frac{|eB\pm eb|^{\frac{3}{2}}}{2^{\frac{1}{2}}\pi}\zeta\left(-\frac{1}{2},\frac{m^2}{2|eB\pm eb|}\right),
\end{eqnarray}
where we have used that the Landau degeneracy per unit area is $|eB\pm eb|/(2 \pi)$. In order to calculate the purely magnetic field effect, we need to subtract the zero-field part. Thus, the one-loop effective Lagrangian density is\footnote{This approach is equivalent to compute an infinite series of one–loop diagrams with the insertion of one, two, \dots external lines (see for instance Ref. \cite{Dittrich1985}).}
\begin{align}\nn
\mathcal{L}^{(1)}_{\pm}&=-(\mathcal{E}^{\pm}_{\text{vac}}(B,b)-\mathcal{E}^{\pm}_{\text{vac}}(0,0))\\
&=-\frac{|eB\pm eb|}{4 \pi}|m|+\frac{|eB\pm eb|^{\frac{3}{2}}}{2^{\frac{1}{2}}\pi}\zeta\left(-\frac{1}{2},\frac{m^2}{2|eB\pm eb|}\right)+\frac{|m|^3}{6 \pi}.
\end{align}
\begin{figure*}
\center
\includegraphics[width=16.4cm]{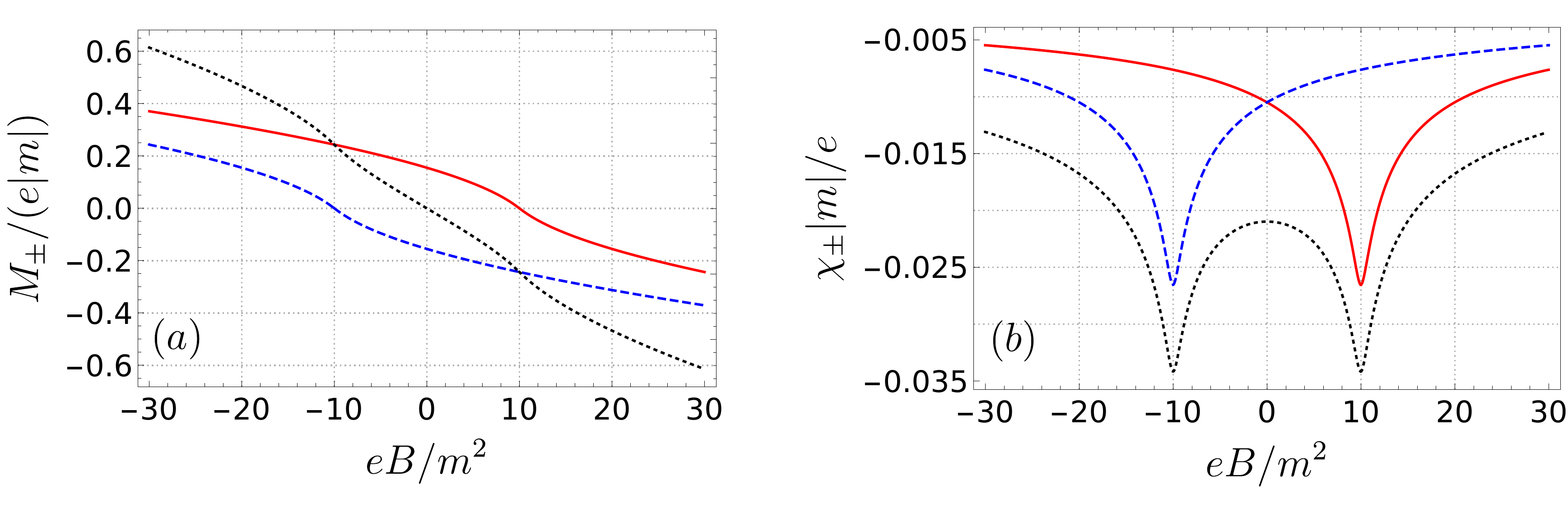}
\caption{(color online) (a) Magnetization vs the external magnetic field, $M_{+}$ (continuous line), $M_{-}$ (dashed line) and the total magnetization $M=M_{+}+M_{-}$ (dotted line). The magnetization ($M_{+}$) in the valley $\vec{K}$ is zero in $eB/m^2=-10$ while the magnetization ($M_{-}$) in the valley $\vec{K}'$ is zero in $eB/m^2=10$. (b) Magnetic susceptibility vs the external magnetic field, $\chi_{+}$ (continuous line), $\chi_{-}$ (dashed line) and the total magnetization $\chi=\chi_{+}+\chi_{-}$ (dotted line). Here we taken $eb/m^2=10$.}
\label{magnetizationfig}
\end{figure*}
Using Eq. (\ref{integralrepre}), we can rewritten the one-loop effective Lagrangian in an integral representation such as
\begin{eqnarray}\nn
\mathcal{L}^{(1)}_{\pm}&=&-\int_0^{\infty}\frac{dt e^{-m^2 t}t^{-\frac{5}{2}}}{8\pi^{3/2}} \left(|eB\pm eb|t\coth (|eB\pm eb|t)-1\right).\\
\end{eqnarray}
For $b=0$, the result is in agreement with what was found in Ref. \cite{Redlich1984a,Andersen1995a,Dittrich1997,Ayala2010c}.
In particular, when $m\rightarrow 0$, we arrive to
\begin{align}
\mathcal{L}^{(1)}_{\pm,m=0}=\frac{|eB\pm eb|^{\frac{3}{2}}}{2^{\frac{1}{2}}\pi}\zeta\left(-\frac{1}{2}\right).
\end{align}
Here $\zeta(x)$ is the Riemann-zeta function. One can compute the orbital magnetization for each valley ($M_{\pm}$) employing the one-loop effective Lagrangian, namely $M_{\pm}=\frac{\partial \mathcal{L}^{(1)}_{\pm}}{\partial B}$. A straightforward calculation gives
\begin{eqnarray}\nn
M_{\pm}&=&-\frac{e}{8\pi}\int_0^{\infty}\frac{dt}{\pi^{1/2}} e^{-m^2 t}t^{-\frac{3}{2}}\left(\coth (|eB\pm eb|t)\right. \\
&&\left. -\frac{|eB\pm eb|t}{\sinh ^2(|eB\pm eb|t)}\right)\text{sgn}(eB\pm eb),
\end{eqnarray}
which can also be written as
\begin{align}\nn
M_{\pm}=&\left[-\frac{e|m|}{4 \pi}-\frac{em^2}{\sqrt{32}\pi|eB\pm eb|^{1/2}}\zeta\left(\frac{1}{2},\frac{m^2}{2|eB\pm eb|}\right)\right.\\
&\left.+3e\frac{|eB\pm eb|^{\frac{1}{2}}}{\sqrt{8}\pi}\zeta\left(-\frac{1}{2},\frac{m^2}{2|eB\pm eb|}\right)\right]\text{sgn}(eB\pm eb).
\end{align}
Therefore, the orbital magnetization displays nonlinear behavior in the magnetic and pseudomagnetic fields (see Fig. \ref{magnetizationfig}a). For $b=0$, the result is in agreement with Refs. \cite{Andersen1995a,Ayala2010c,Slizovskiy2012}  and in the limit $m\rightarrow 0$ the magnetization is
\begin{align}\label{magnezerostronglimit}
M_{\pm}=3e\frac{|eB\pm eb|^{\frac{1}{2}}}{\sqrt{8}\pi}\zeta\left(-\frac{1}{2}\right)\text{sgn}(eB\pm eb),
\end{align}
for each valley\footnote{It should be noted that Eq. (\ref{magnezerostronglimit}) also corresponds to the dominant term in the strong field expansion.}. Notably, the value and sign of magnetization are different for each valley. Thus, they can be modified by the strains or varying the applied magnetic field. In particular, the magnetization of one valley could be zero while the other does not, as it can be seen in Fig. \ref{magnetizationfig}a.
\par
Having calculated $M_{\pm}$, one can compute the magnetic susceptibility for each valley in the presence of magnetic and pseudomagnetic fields, which is simply given by $\chi_{\pm}=\frac{\partial M_{\pm}}{\partial B}$, thus
\begin{eqnarray}\nn
\chi_{\pm}&=&\frac{e}{16\sqrt{2} \pi |eB\pm eb|^{\frac{5}{2}}}\left[ 12 |eB\pm eb|^2\zeta\left(-\frac{1}{2},\frac{m^2}{2|eB\pm eb|}\right)\right.\\\nn
&&\left. -4m^2|eB\pm eb|\zeta\left(\frac{1}{2},\frac{m^2}{2|eB\pm eb|}\right)-m^4\zeta\left(\frac{3}{2},\frac{m^2}{2|eB\pm eb|}\right)\right]\\
\end{eqnarray}
In the presence of magnetic and pseudomagnetic fields, the total susceptibility has two minimums, see Fig. \ref{magnetizationfig}b, which is a distinctive feature compared with the case $eb=0$ \cite{Slizovskiy2012}. Finally, in the limit $m\rightarrow 0$ the susceptibility is
\begin{align}
\chi_{\pm}&=\frac{3 e \zeta(-1/2)}{\sqrt{32}\pi|eB\pm eb|^{1/2}},
\end{align}
which is divergent when $B=\pm b$. Since $\chi_{\pm}(B\pm b=0,m\neq 0)$ is finite, it can be concluded that the mass acts as a regulator.
\section{Induced charge density}\label{secIndChar}
In this section, we derive an expression for the induced charge density in the presence of uniform real and pseudomagnetic fields. We first noticed, that as in the case of the fermionic condensate, the formula $-\text{tr} \gamma^{\mu} S(x,x)=\langle 0| j^{\mu}|0\rangle$ also deserves to be revised. The current must be understood as
\begin{eqnarray}\nn
j^{\mu}\rightarrow :j^{\mu}(x)&:=&:e\bar{\Psi}(x)\gamma^{\mu}\Psi(x):\\\nn
&=&\frac{e}{2}[\bar{\Psi}(x),\gamma^{\mu}\Psi(x)]\\
&=&\frac{1}{2}\gamma^{\mu}_{\dot{\alpha}\alpha}[\bar{\Psi}_{\dot{\alpha}}(x),\Psi_{\alpha}(x)],
\end{eqnarray}
which is the correct definition of the current operator, where it has been subtracted an infinite charge of the vacuum state \cite{Dittrich1985}. In an analogous way to what was done before, we can find that in the Furry picture the vacuum expectation value of the current operator is
\begin{eqnarray}\nn
&&\langle 0|:j_i^{\mu}(x):|0\rangle =\frac{e}{2}\langle 0|[\bar{\Psi}_i(x),\gamma^{\mu}\Psi_i(x)]|0\rangle\\\nn
&&=\frac{e}{2}\SumInt_{n}
\SumInt_p \left( \bar{\psi}^{(-)}_{i,n,p}(x)\gamma^{\mu}\psi^{(-)}_{i,n,p}(x)-\bar{\psi}^{(+)}_{i,n,p}(x)\gamma^{\mu}\psi^{(+)}_{i,n,p}(x)\right).\\
\end{eqnarray}
For constant real and pseudomagnetic fields, using the orthogonality of Hermite polynomials, one can show that
\begin{align}\label{InduCurrendenpm}
\langle 0|:j_{\pm}^{i}(x):|0\rangle &=0, \ \ \ (i=1,2),
\end{align}
\textit{i.e} the induced current vanishes. A nonvanishing vacuum current would arise in the presence of an external electric field \cite{Andersen1995a}. On the other hand,
the induced charge density is given by
\begin{align}\label{InduChargedenpm}
\langle 0|\rho(x)_{\pm}|0\rangle=\langle 0|:j_{\pm}^{0}(x):|0\rangle &=\pm\frac{\text{sgn}(m)}{4\pi}e^2(B\pm b).
\end{align}
Curiously, in contrast to $\mu^{\pm}$, the induced charge density only receives contributions from the lowest Landau level (LLL), even if $m\neq 0$. An alternative technique to compute the charge density is through spectral function. It can be shown that the induced charge is \cite{Reuter1986a,Dittrich1986L}
\begin{align}
Q_{\pm}=\int d^2x\langle 0|\rho(x)_{\pm}|0\rangle=-\frac{e}{2}\lim_{\substack{s\rightarrow 0^{+}}}\eta(\mathcal{H}_{\pm},s),
\end{align}
where
\begin{align}
\eta(\mathcal{H}_{\pm},s)=\sum_n|E_{n}^{\pm}|^{-s}\text{sgn}(E_{n}^{\pm}).
\end{align}
is the $\eta$ invariant of Atiyah, Patodi, and Singer. Using the Landau degeneracy per unit area and noting that except for the LLL, for each eigenvalue $E_n^{\pm}$ there is an eigenvalue $-E_n^{\pm}$, then we obtain Eq. (\ref{InduChargedenpm}) as we should. Therefore, the total induced charge density is
\begin{eqnarray}\nn
\langle 0|\rho(x)|0\rangle&=&\langle 0|\rho(x)_{+}|0\rangle+\langle 0|\rho(x)_{-}|0\rangle\\\nn
&=&\frac{\text{sgn}(m)}{4\pi}e^2(B+b)-\frac{\text{sgn}(m)}{4\pi}e^2(B-b)\\\label{cchargepseudo}
&=&\frac{\text{sgn}(m)}{2\pi}e^2b,
\end{eqnarray}
which is observable (and measurable) when taking a nonzero pseudomagnetic field. This is remarkable since 
this is not possible in the case of a pure magnetic field \cite{Semenoff1984a}. Eq. (\ref{cchargepseudo}) shows that in the presence of pseudomagnetic fields, the system has a parity anomaly, even with two fermionic species. Eqs. (\ref{InduCurrendenpm}) and (\ref{InduChargedenpm}) are related with the Chern-Simons relation, which in the case of zero pseudomagnetic field reads \cite{Semenoff1984a}
\begin{align}
\langle 0|:j_{\pm}^{\mu}(x):|0\rangle &=\pm\frac{e^2}{8\pi}\text{sgn}(m)\epsilon^{\mu\nu \lambda}F_{\nu\lambda}.
\end{align}
It is clear from this equation equation that when the two representations are present, the vacuum expectation value of the total current is always zero.
\section{Conclusions}\label{conclutionsfinal}
In summary, we have examined the Dirac Hamiltonian in $2+1$ dimensions and found an infinite number of non-equivalent realizations of parity, charge conjugation, and time-reversal transformations for the reducible representation. We have then explored how the interplay between real and pseudomagnetic fields affects some aspects of the three-dimensional quantum field theory. For the case of uniform magnetic and pseudomagnetic fields, by employing a non-perturbative approach we have found that: (i) The $c-$condensate is the appropriate order parameter for studying the breaking of chiral, parity and time-reversal symmetries. (ii) One can control the magnetization, susceptibility, and the dynamical mass independently for each valley by straining and varying the applied magnetic field.  (iii) The dynamical mass generated is due to two terms, the usual mass term ($m\bar{\psi}\psi$) and a Haldane mass term ($m_{\tau}\bar{\psi}i\Gamma^{35}\psi$), being the latter, for the case in which the two fields are simultaneously different from zero, the one that breaks parity and time-reversal symmetries. (iv) For non-zero pseudomagnetic field, the total induced ``vacuum'' charge density is not null. This last result implies that strained single-layer graphene exhibits a parity anomaly. Therefore, strained graphene in the presence of an external magnetic field has distinctive features compared with QED$_{2+1}$, which lacks the aforementioned consequences (i)-(iv). Finally, it would be interesting to extend our calculations to include the effect of Coulomb interactions on the magnetic catalysis, symmetry breaking, dynamical mass generation, etc. See, for instance, the study of the magnetic catalysis in unstrained graphene in the weak-coupling limit \cite{Semenoff2011a}.
\begin{acknowledgements}
J. A. S\'{a}nchez is grateful to F. T. Brandt, C. M. Acosta and J. S. Cort\'{e}s for useful comments.
\end{acknowledgements}
\appendix
\section{Exact solutions of the Dirac equation in the presence of uniform real and pseudomagnetic fields}	\label{ap:ConsRPMF}
In the presence of a constant real ($B$) and pseudo ($b$) magnetic
fields, one can choose the real potential and the pseudo-potential as $\vec{A}=(0,Bx)$ and $\vec{a}^{35}=(0,bx)$, respectively. For the valley $\vec{K}$ and with $m\geq0$, it is straightforward to find the spectrum and the solutions of
Eq. (\ref{DiracEpm}). However, we need to find them independently for the
cases $e(B-b)>0$ and $e(B-b)<0$. For the first case $e(B-b)>0$, the spectrum and solutions are given by
\begin{equation}\label{EnergyAB+}
E_n^{-}=\pm\sqrt{m^2+2|eB-eb|n},
\end{equation}
\begin{equation}\label{NESABP+0}
\psi^{(+)}_K=N^{-}_ne^{-i|E^{-}_n|t+ipy}\left( \begin{array}{c}
(|E^{-}_n|+m)I^{-}(n,p,x)\\
-\sqrt{2|eB-eb|n}I^{-}(n-1,p,x)\\
\end{array} \right),
\end{equation}
\begin{equation}\label{NESABN+0}
\psi^{(-)}_K=N^{-}_ne^{i|E^{-}_n|t-ipy}\left( \begin{array}{c}
\sqrt{2|eB-eb|n}I^{-}(n,-p,x)\\
(|E^{-}_n|+m)I^{-}(n-1,-p,x)\\
\end{array} \right),
\end{equation}
where
\begin{equation}
N^{\pm}_n=\frac{1}{\sqrt{2|E^{\pm}_n|(|E^{\pm}_n|+m)}},
\end{equation}
\begin{eqnarray}\nn
I^{\pm}(n,x,p)&=&\frac{|eB\pm eb|^{\frac{1}{4}}}{\sqrt{2^n n!}\pi^{\frac{1}{4}}}
H_n\left[\sqrt{|eB\pm eb|}\left(x-\frac{p}{|eB\pm eb|}\right)\right]\\
&&\times \exp\left(-\frac{|eB\pm eb|}{2}\left(x-\frac{p}{|eB\pm
eb|}\right)^2\right)
\end{eqnarray}
with $I(n=-1,p,x)=0$ and $E_n^{+}=\pm\sqrt{m^2+2|eB+eb|n}$.
Note that the lowest Landau level (LLL) describes
particle (fermion) states with energy $E_0=m\geq 0$. In order to find the solutions for the case $e(B-b)<0$, one can use the charge-conjugate operator, Eq. (\ref{Ccharge2x2}), $\mathcal{C}\psi=-\gamma^2(\bar{\psi})^T=\sigma_1\psi^*$ \cite{Ahmed1985a}, so, the solutions are given by
\begin{equation}\label{NESABP-0}
\psi^{(+)}_K=-N^{-}_ne^{-i|E^{-}_n|t+ipy}\left( \begin{array}{c}
(|E^{-}_n|+m)I^{-}(n-1,p,x)\\
\sqrt{2|eB-eb|n}I^{-}(n,p,x)\\
\end{array} \right),
\end{equation}
\begin{equation}\label{NESABN-0}
\psi^{(-)}_K=-N^{-}_ne^{i|E^{-}_n|t-ipy}\left( \begin{array}{c}
-\sqrt{2|eB-eb|n}I^{-}(n-1,-p,x)\\
(|E^{-}_n|+m)I^{-}(n,-p,x)\\
\end{array} \right),
\end{equation}
Now the LLL describes hole (antifermion) states.
\par
One can find the solution for the valley $\vec{K}'$ noting that the change of representation is equivalent to change $m\rightarrow -m$ and $B-b\rightarrow B+b$ in the solutions (\ref{NESABP+0}), (\ref{NESABN+0}), (\ref{NESABP-0}) and (\ref{NESABN-0}). For the point $\vec{K}'$, the LLL describes a hole (antifermion) state for $e(B+b)>0$ and a particle (fermion) state for $e(B+b)<0$. Hence, we can have a fermion (antifermion) state in the LLL simultaneously to $\vec{K}$ and $\vec{K}'$ if the condition $eb<eB<-eb$ ($eb>eB>-eb$) is satisfied.
\section{Magnetic condensate}\label{ap:MagConApen}
In this Appendix, we will compare the calculation of fermionic condensate in the Furry picture with the method via fermion propagator. We then demonstrate that the method via the fermion propagator does not calculate the vacuum expectation value of the product of two field operators (magnetic condensate), but rather a vacuum expectation value of the commutator of two field operators. Finally, we show that the commutator is the order parameter is for chiral (or parity) symmetry breaking, instead of the magnetic condensate, as it is often asserted in the literature.
\subsection{Magnetic condensate via furry picture}\label{ap:MagConFurryP}
Let us consider the fermionic condensate via the Furry picture in a constant background magnetic field. The explicit solution can be found in the Appendix (\ref{ap:ConsRPMF}), taking $b=0$. Inserting Eq. (\ref{NESABN+0}) into Eq. (\ref{psiAPsiAG}), we obtain that for the irreducible representation $\mathcal{R}_1$ (valley $\vec{K}$), for $eB>0$ and $m>0$, the fermion condensate  is
\be\nn
\langle 0| \bar{\Psi}_K\Psi_K|0 \rangle&=&
\sum_{n=0}^{\infty}\int dp \frac{N^{2}_n}{2\pi} \left[ 2|eB|nI(n,-p,x)^2.\right. \\\nn
&&\left.-(|E_n|+m)^2I(n-1,-p,x)^2 \right]\\\label{CONDAB+0}
&=&-\frac{|eB|}{2\pi}
\sum_{n=1}^{\infty} \frac{m}{|E_n|},
\ee
with $|E_n|=\sqrt{m^2+2e|B|n}$ and where we used that
\be
\int dp I(n-1,p,x)^2=\left\{ \begin{array}{cc}
0 & \text{if } n=0,\\
|eB| & \text{if } n>0,\\
\end{array} \right.
\ee
and that for $n=0$ all the terms are zero. Now, for $eB<0$,
inserting Eq. (\ref{NESABN-0}) into Eq. (\ref{psiAPsiAG}) yields
\begin{eqnarray}\nn
\langle 0| \bar{\Psi}_K\Psi_K|0 \rangle=
&=&\sum_{n=0}^{\infty}\int dp \frac{N^{2}_n}{2\pi} \left[2|eB|nI(n-1,-p,x)^2\right. \\\nn
&&\left.-(|E_n|+m)^2I(n,-p,x)^2 \right]\\\label{CONDAB-0}
&=&-\frac{|eB|}{2\pi}-\frac{|eB|}{2\pi}\sum_{n=1}^{\infty} \frac{m}{|E_n|}.
\end{eqnarray}
It is clear now that $n=0$ contribute to the condensate. In a similar way, the condensates in the irreducible representation $\mathcal{R}_2$ (valley $\vec{K}'$) are
\begin{align}\label{CONDCB+0}
\langle 0|
\bar{\Psi}_{K'}\Psi_{K'}|0
\rangle &=-\frac{|eB|}{2\pi}-\frac{|eB|}{2\pi}\sum_{n=1}^{\infty}\frac{m}{|E_n|},\\\label{CONDCB-0}
\langle 0|\bar{\Psi}_{K'}\Psi_{K'}|0
\rangle &=-\frac{|eB|}{2\pi}\sum_{n=1}^{\infty}\frac{m}{|E_n|},
\end{align}
for $eB>0$ and $eB<0$, respectively. In the fermion condensates, the term $\sum_{n=1}^{\infty}\frac{1}{|E_n|}$ is in general divergent. However, the results are understood by means of an appropriate analytic continuation.
\par
Eqs. (\ref{CONDAB+0}) and (\ref{CONDCB+0}) are in agreement with what was found in Ref. \cite{Cea2000}. When $m<0$, one just need to exchange results in Eq. (\ref{CONDAB+0}) with the one of Eq. (\ref{CONDAB-0}), and Eq. (\ref{CONDCB+0}) with Eq. (\ref{CONDCB-0}). In the limit $m\rightarrow 0^{+}$, $\it i.e.$ for massless fermions, the fermionic condensates are
\be\label{m0BA} \langle 0|
\bar{\Psi}_{K}\Psi_{K}|0 \rangle= \left\{ \begin{array}{cc}
-|eB|/(2\pi) & \text{if } eB<0,\\
0 & \text{if } eB>0,\\
\end{array} \right.
\ee
and
\be\label{m0BB}
\langle 0| \bar{\Psi}_{K'}\Psi_{K'}|0 \rangle=
\left\{ \begin{array}{cc}
0 & \text{if } eB<0,\\
-|eB|/(2\pi) & \text{if } eB>0.\\
\end{array} \right.
\ee
Notably, only if the LLL has negative energy states there is a non-vanished magnetic condensate. The condensate in the $4\times 4$ reducible representation is simply the sum of the irreducible representations
\be\label{m0b4x4}
\langle 0|  \bar{\Psi}\Psi|0\rangle=\langle 0|  \bar{\Psi}_{K}\Psi_{K}|0\rangle+\langle 0|  \bar{\Psi}_{K'}\Psi_{K'}|0\rangle=-\frac{|eB|}{2\pi},
\ee
which is in agreement with the results of Refs. \cite{Dunne1996,Das1996,Das1997,Farakos1998b,Anguiano2007,Cea2000}.
\subsection{Magnetic condensate via the fermion propagator}
Now we consider the usual way to calculate the vacuum condensate
using the fermion propagator \cite{Gusynin1994}
\be\label{propagator}
S_F(x,y)=\langle 0|T[\Psi(x)\bar{\Psi}(y)]|0\rangle,
\ee
where $T$ is the time-ordering operator
\be\nn
\langle 0|T[\Psi(x)\bar{\Psi}(y)]|0\rangle&=&\theta(x_0-y_0)\langle 0|\Psi(x)\bar{\Psi}(y)|0\rangle\\\label{PropaD}
&&
-\theta(y_0-x_0)\langle 0|\bar{\Psi}(y)\Psi(x)|0\rangle,
\ee
with $\theta(x)$ the Heaviside step function. In the presence of a magnetic field, $S_F(x,y)$ can be calculated by using the Schwinger (proper time) approach \cite{Schwinger1951}
\be
S_F(x,y)=\exp\left[\int_x^y dx_{\mu}A_{ext}^{\mu}\right]\tilde{S}(x-y),
\ee
where the integral is calculated along the straight line, and
the Fourier transform of $S_F(x)$ (in Euclidean space) is ($k_3=-ik_0$)
\be\nn
\tilde{S}(k)&=&-i\int_{0}^{\infty}ds\exp\left[-s\left(m^2+k^2_3+\mathbf{k}^2\frac{\tanh(eBs)}{eBs}\right)\right]\\\nn
&&\times \left(-k_{\mu}\gamma_{\mu}+m\mathbf{1}-i(k_2\gamma_2-k_1\gamma_2)\tanh(eBs)\right)\\
&&\times\left(1-i\gamma_1\gamma_2\tanh(eBs)\right),
\ee
with $\mathbf{1}$ and $\gamma^{\mu}$ the $4\times 4$ ($2\times2$) identity matrix
and the $4\times 4$ ($2\times2$) $\gamma$-matrices in the reducible (irreducible) representations. The magnetic condensate has been calculated as \cite{Gusynin1994,Gusynin1995a,Gusynin1996a}
\be\label{m0b4x4S}
\langle 0| \bar{\Psi}(x)\Psi(x)|0 \rangle^S=-\lim_{x\rightarrow y} \text{tr} S_F(x,y),
\ee
where the superscript $S$ emphasizes that the condensate is computed via the propagator. For the reducible representations, in the limit $m\rightarrow 0$, the condensate reads
\cite{Gusynin1994}
\be
\langle0| \bar{\Psi}(x)\Psi(x)|0 \rangle^S=-\frac{|eB|}{2\pi},
\ee
which is in agreement with Eq. (\ref{m0b4x4}). However,
for the irreducible representations the magnetic condensate is
\be\label{m0ABSF}
\langle0| \bar{\Psi}_{K(K')}(x)\Psi_{K(K')}(x)|0 \rangle^S=-\frac{|eB|}{4\pi},
\ee
which is the same for both representations, but it differs from what was obtained previously in Eqs. (\ref{m0BA}) and (\ref{m0BB}). To explain why the condensates are different, let us critically review the calculation through propagator\footnote{
Using the so-called Ritus eigenfunctions method, it has recently been found Eq. (\ref{m0ABSF}) \cite{Raya2010}, however, this method also uses the propagator to calculate the condensate.}. If one part of the definition of the propagator Eqs. (\ref{propagator})
and (\ref{PropaD}), it is clear that $\lim_{x\rightarrow y} \text{tr} S(x,y)$ depends on how one takes the limit, namely
\be
\text{tr} \lim_{\substack{x_0\rightarrow y_0^{+}\\ \mathbf{x}\rightarrow \mathbf{y}}} S_F(x,y)&=&\langle 0|\Psi_{i}(x) \bar{\Psi}_{i}(x)|0\rangle^S, \\
\text{tr} \lim_{\substack{x_0\rightarrow y_0^{-}\\ \mathbf{x}\rightarrow \mathbf{y}}} S_F(x,y)
&=&-\langle 0|\bar{\Psi}_{i}(x)\Psi_{i}(x)|0\rangle^S.
\ee
Using Eq. (\ref{PsiAGpsiA}) one can show that, in the limit $m\rightarrow 0^{+}$, the fermion condensates $\langle \Psi_{i}\bar{\Psi}_{i}\rangle$ for the irreducible representations are
\be\label{m0BAP}
\langle 0|
\Psi_{K}(x)\bar{\Psi}_{K}(x)|0 \rangle&=&
\left\{ \begin{array}{cc}
0 & \text{if } eB<0,\\
|eB|/(2\pi) & \text{if } eB>0,\\
\end{array} \right.\\
\label{m0BBP}
\langle 0|
\Psi_{K'}(x)\bar{\Psi}_{K'}(x)|0 \rangle&=& \left\{ \begin{array}{cc}
|eB|/(2\pi) & \text{if } eB<0,\\
0 & \text{if } eB>0,\\
\end{array} \right.
\ee
and
\be\nn
\langle0| \Psi(x)\bar{\Psi}(x)|0 \rangle&=&\langle 0|
\Psi_{K}(x)\bar{\Psi}_{K}(x)|0 \rangle+\langle 0|\Psi_{K'}(x)\bar{\Psi}_{K'}(x)|0 \rangle\\\label{m0b4x4P}
&&=\frac{|eB|}{2\pi},
\ee
for the reducible representation. It is clear that for irreducible representations neither $\langle \bar{\Psi}_{K(K')}\Psi_{K(K')} \rangle$ nor $\langle \Psi_{K(K')}\bar{\Psi}_{K(K')}\rangle$ coincide with $\langle\bar{\Psi}_{K(K')}\bar{\Psi}_{K(K')} \rangle^S$. The calculation of the condensate in Refs. \cite{Gusynin1994,Gusynin1995a} was actually calculated
by taking $x=y$. This will formally lead to calculate the propagator in $x_0=y_0$,
which depends on $\theta(0)$. In the literature there is not a consensus of what should be the value of $\theta(0)$. For instance, $\theta(0)$ can be $1$, $0$ or $1/2$ (page 24 in \cite{Zee2010}). However, the only consistent value with Eqs. (\ref{m0BA}), (\ref{m0BB}), (\ref{m0b4x4}), (\ref{m0BAP}), (\ref{m0BBP}) and (\ref{m0b4x4P}) is the value of $1/2$,
since
\be\nn
\text{tr} S_F(x,x)&=&\frac{\langle 0|\Psi(x)\bar{\Psi}(x)|0\rangle}{2}-\frac{\langle 0|\bar{\Psi}(x) \Psi(x)|0\rangle}{2}\\\label{SchwingerPres}
&=&\frac{1}{2}\langle 0|[\Psi(x),\bar{\Psi}(x)]|0\rangle
\ee
so,
\be
\langle 0|\bar{\Psi}(x) \Psi(x)|0\rangle=\langle 0|\Psi(x)\bar{\Psi}(x)|0\rangle-2\text{tr} S_F(x,x),
\ee
which gives us the correct value for all condensates. In fact this last choice was the Schwinger's choice: \textit{``the average of the forms obtained by letting $y$ approach $x$ from the future, and from the past''} \cite{Schwinger1951}. Therefore, the Heaviside step function in the fermion propagator should be understood as
\be
\theta(t)=
\left\{
    \begin{array}{ll}
        0  & \text{if } t < 0 \\
        1/2  & \text{if } t = 0 \\
        1 & \text{if } t > 0.
    \end{array}
\right.
\ee
In summary, $-\text{tr} S_F(x,x)\neq\langle 0|\bar{\Psi}(x) \Psi(x)|0\rangle$.
We believe that this has not been noticed before since, in particular, when the condensate is computed in $2+1$ dimensions in a reducible representation, the two values match, which as we showed is just a coincidence\footnote{One can show that in $3+1$ dimensions the two values also match, see Ref. \cite{Gusynin1995c} for the calculation via fermion propagator and reference \cite{Anguiano2007} for the calculation in the Furry picture.}.
\subsection{Order parameter}\label{OrderParamAppen}
It is widely claimed in the literature that in the reducible representations the fermion condensate $\langle \bar{\Psi}\Psi\rangle$ is the order parameter of dynamical chiral symmetry breaking \cite{Gusynin1994,Gusynin1995a,Gusynin1995b,Dunne1996,Das1996}. However, as we have just showed, a part of the literature calculates $\langle \bar{\Psi} \Psi\rangle$ and another part calculates $\langle [\bar{\Psi},\Psi]\rangle/2$ as the order parameter, always assuming that $\langle \bar{\Psi} \Psi\rangle$ is being calculated. Thus, we may ask what is the order parameter for chiral (or parity) symmetry breaking. Since in reducible representations the two values match, we will use the irreducible representations to solve this puzzle.
\par
Although in irreducible representations we have not a chiral symmetry, as we mentioned above, the mass term breaks parity and time-reversal symmetry. If the fermion condensate is the order parameter of dynamical parity and time-reversal symmetry breaking, Eqs. (\ref{m0BA}) and (\ref{m0BB}) would lead us to conclude that these broken symmetries depend on the magnetic field orientation. Moreover, since the fermionic condensate is zero for $eB>0$ ($eB<0$) in the representation $\mathcal{R}_1$ ($\mathcal{R}_2$) there would not be a dynamical symmetry breaking in this case and therefore would not have a magnetically induced mass. However, if the order parameter is the commutator, which is independent of the orientation of the magnetic field and the representation, we would obtain a magnetically induced mass for $eB\neq 0$.
\par
In QED$_{2+1}$ and RQED$_{4,3}$ the dynamical mass generation has already been investigated \cite{Hoshino1989a,Gusynin1994,Shpagin1996,Farakos1998a,Farakos1998b,Raya2010,khalilov2019a}. In particular, the dynamical mass generation with a two-component fermion was examined in \cite{Hoshino1989a,Raya2010,khalilov2019a}. From the results of these works, it can be concluded that there is no evidence that the dynamical mass depends on the orientation of the magnetic field. Moreover, under some assumptions, an explicit formula was found for the dynamical mass when it is much smaller than the magnetic field \cite{Raya2010}. One can show that the dynamic mass found is independent of the orientation and the representation of the $\gamma$-matrices. Therefore, the order parameter for parity (or chiral) symmetry breaking is not the vacuum expectation value of the product of two field operators (magnetic condensate), but rather a vacuum expectation value of the commutator of two field operators. The latter is equivalent to the trace of the fermion propagator evaluated at equal space-time points.


\end{document}